\documentclass[10pt,a4paper]{article}
\usepackage[utf8]{inputenc}
\usepackage[english]{babel}
\usepackage{amsmath}
\usepackage{amsthm}
\usepackage[hidelinks,colorlinks,allcolors=blue]{hyperref}
\usepackage{amsfonts}
\usepackage{amssymb}
\usepackage{graphicx}
\usepackage{float}
\usepackage{enumerate}
\usepackage{bbm}
\usepackage{xcolor}
\usepackage{hhline}
\usepackage[left=2cm,right=2cm,top=2cm,bottom=2cm]{geometry}
\usepackage[round]{natbib}   
\usepackage{authblk}
\title{Bivariate modelling of precipitation and temperature using a non-homogeneous hidden Markov model}
\date{}

\author[1,2]{Augustin Touron }
\author[2]{Thi-Thu-Huong Hoang}
\author[2]{Sylvie Parey}
\affil[1]{Laboratoire de Math\'ematiques d'Orsay, Universit\'e Paris-Sud}
\affil[2]{EDF R\&D}

\usepackage{tikz} 

\usetikzlibrary{arrows,decorations.pathmorphing,backgrounds,positioning,fit,%
automata,calc} 
\tikzset{
  main/.style={circle, minimum size = 10mm, thick, draw =black!80, node distance = 7mm},
  main2/.style={minimum size = 10mm},
  connect/.style={-latex, thick},
  box/.style={rectangle, draw=black!100}
}

\newcommand{\R}{\mathbb{R}}

\DeclareMathOperator*{\argmax}{arg\,max}
\DeclareMathOperator*{\argmin}{arg\,min}

\begin{document}

\maketitle

\begin{abstract}
Aiming to generate realistic synthetic times series of the bivariate process of daily mean temperature and precipitations, we introduce a non-homogeneous hidden Markov model. The non-homogeneity lies in periodic transition probabilities between the hidden states, and time-dependent emission distributions. This enables the model to account for the non-stationary behaviour of weather variables. By carefully choosing the emission distributions, it is also possible to model the dependance structure between the two variables. The model is applied to several weather stations in Europe with various climates, and we show that it is able to simulate realistic bivariate time series.
\end{abstract}

\section{Introduction}

Historically, the management and planning of electricity demand and generation has involved long lasting observed or synthetic temperature time series, because temperature is the main driver of electricity demand. Then the centralized generation facilities are managed to match the anticipated demand. With a growing part of less manageable renewable generation based on wind and solar, the need for the same type of meteorological information, but not restricted to temperature anymore, emerges. The necessity for the system to be robust to as many different meteorological situations as possible involves a need for large samples of consistent evolutions of many meteorological variables, such as temperature, wind speed, solar radiation and rainfall for example. Since observation or reanalysis products are all available over quite limited time periods, stochastic weather generators are valuable tools to enrich the samples. For example, stochastic generators for temperature are commonly used as part of pricing derivatives, in relation with energy prices (\cite{campbell2005},  \cite{mraoua2007}, \cite{benth2011}).\\

\noindent Single site multivariate models have been studied for several decades. The most widely cited model for weather variables has been proposed by Richardson \citep{richardson81} in the framework of crop development, and lots of models have then been developed on the same basis (see \cite{wilks1999} for a review). These models condition the evolution of the non-precipitation variables on two states based on occurrence and nonoccurrence of rainfall. Then the simulation of the non-precipitation variables is obtained through a multivariate autoregressive process, mostly using Gaussian distributions. In some cases, the autoregressive parameters depend on weather types. \cite{flecher2010} extend this concept by using more weather types and skew normal distributions. The weather types are identified through classifications of the rainy and non-rainy days separately for each season and the number of weather types is chosen according to the BIC criterion. \cite{vrac2007} define a model used for precipitation downscaling based on weather types identified a priori through classifications either of the precipitation data or of exogenous atmospheric variables. However, such a priori definitions of the weather types may not be optimal to infer the stochastic properties of the variable to generate.\\

\noindent Hidden Markov Models (HMM) introduce the weather types as latent variables. In theses models, the states form a latent Markov chain and conditionally to the states, the observations are independent. Although simple, they are very flexible:
\begin{itemize}
\item the determination of the states is data driven instead of depending on arbitrarily chosen exogeneous variables,
\item they allow non-parametric state-dependent distributions,
\item using few parameters, they are able to model complex time dependence for the observations. 
\end{itemize}

\noindent Homogeneous HMMs are generally used for multisite generation either of rainfall occurrences \citep{zucchini91} or of the whole rainfall field. \cite{kirshner2005} proposes an overview and tests different options for the multivariate emissions, from conditional independence to complex dependence structures, going through tree structures. \cite{ailliot2015} offers a more recent overview of the weather type based stochastic weather generators, including HMMs. Extensions to Non Homogenous HMMs are also proposed in order to introduce a diurnal cycle \citep{ailliot2012} or to let the probability of a hidden state depend on the value of an external input variable (\cite{hughes1994}, \cite{hughes1999}).\\

\noindent Recently, new ways of generating meteorological variable have been studied. As an example, \cite{peleg2017} designed a model mixing physically and stochastically based features in order to generate gridded climate variables at high spatial and temporal resolution.

\paragraph{Our contribution}In this paper, we introduce a non homogeneous HMM for the single site generation of temperature and rainfall at different locations in Europe presenting different climatic conditions. The model is here designed for a single site generation, because electricity load and generation balance is more and more studied at a very local scale in relation to the decentralization of electricity generation based on renewables. Furthermore, global balance is generally studied on the basis of geographical (possibly weighted) averages of the demand and generation respectively. The proposed HMM is non homogeneous because the seasonality is introduced in the transition matrix between the hidden states, as well as in the state-dependent distributions. Most of stochastic weather generators in the literature elude the problem of the non-stationarity of weather variables by defining a different model for each season or month independently, assuming local stationarity inside each block. For example, \cite{lennartsson2008} consider blocks of lengths one, two or three months. This approach has several drawbacks:
\begin{itemize}
\item the local stationarity assumption may be difficult to check,
\item the data used to fit each model is obtained as a concatenation of data that do not belong to the same year. This is a problem if our data exhibits a strong time dependence,
\item a stochastic weather generator should be able to simulate long times series using only one model.
\end{itemize}
Our model, in contrast, allows the generation of synthetic climate variables, without splitting the data, and without any pre-processing. Furthermore, the generation of temperature implies handling the warming trend. Whereas \cite{flecher2010} proposed a standardization of the temperature and radiation fields beforehand, the choice has been made here to explicitly introduce a trend in the temperature generation.\\

\noindent One of the main issues when dealing with multivariate modeling is being able to capture the possibly complex dependence structure between the variables. We introduced the state-dependent distributions of our HMM as mixtures of tensor products (see equation \eqref{loi-emission}). This allows us, provided that the number of components in the mixture is large enough and that the marginals are well chosen, to approximate any state-dependent bivariate distribution for temperature and precipitation, without any a-priori on their dependence structure. Besides, it makes it easy to generalize the model to a larger number of variables without changing its global definition.\\

\noindent Although they are not detailed in this paper, we studied the theoretical properties of our model and gave theoretical guarantees for the convergence of the maximum likelihood estimator (see Section \ref{sec2}).

\paragraph{Outline}A general description of our non homogeneous HMM is given in Section \ref{sec2}, as well as a reminder of the existing theoretical results linked to our model. Section \ref{sec3} deals with the application of the model to precipitation and temperature observations. In this section, we present the data used, then we describe more precisely the modeling framework by giving the parametrization of our model that is specific to this application, and we discuss the results for the different locations. We finally go to the main conclusions and perspectives in the last section.

\section{Model}\label{sec2}

In this section, we describe the mathematical framework in which we developed our model. We first give a general definition of non-homogeneous hidden Markov models, before addressing the topic of theoretical results regarding these models. The full details of the parametrization of our model are given in section \ref{modele}, along with its application to climate data.

\subsection{General formulation}\label{modgen}

\begin{figure}
\centering
\begin{tikzpicture}[scale=1]
  \node[box,draw=white!100] (Latent) {\textbf{Observed}};
  \node[main] (L1) [right=of Latent] {$Y_{t-1}$};
  \node[main,minimum size=1.2cm] (L2) [right=of L1] {$Y_t$};
  \node[main] (L3) [right=of L2] {$Y_{t+1}$};
  \node[main2] (Lt) [right=of L3] {$\in \mathsf{Y}$ (polish)};
  \node[main,fill=black!10] (O1) [below=of L1] {$X_{t-1}$};
  \node[main,fill=black!10,minimum size=1.2cm] (O2) [below=of L2] {$X_t$};
  \node[main,fill=black!10] (O3) [below=of L3] {$X_{t+1}$};
  \node[main2] (Ot) [below=of Lt] {$\  \in \mathsf{X}$ (finite)};
  \node[box,draw=white!100,left=of O1] (Observed) {\textbf{Hidden}};
  \path (Observed) -- node[auto=false]{\ldots} (O1);
  \path (O1) edge [connect] (O2)
        (O2) edge [connect] (O3) 
        (O3) -- node[auto=false]{\quad \ldots} (Ot);
  \path (O1) edge [connect] (L1);
  \path (O2) edge [connect] (L2);
  \path (O3) edge [connect] (L3);

  \draw [dashed, shorten >=-1cm, shorten <=-0.5cm]
      ($(Latent)!0.5!(Observed)$) coordinate (a) -- ($(L3)!(a)!(O3)$);
\end{tikzpicture}
\caption{The dynamic of a hidden Markov model}\label{hmm}
\end{figure}
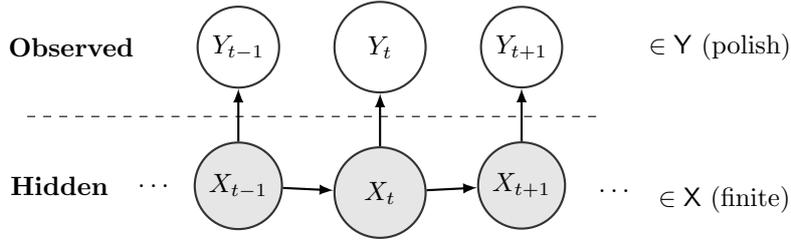

Let us first recall the definition of a finite state space hidden Markov model (HMM). Let $K$ be a positive integer and $\mathsf{X}=\{1,\dots,K\}$. Let $\mathsf{Y}$ be a Polish space equiped with its Borel $\sigma$-algebra $\mathcal{Y}$. A (non-homogeneous) hidden Markov model with state space $\mathsf{X}$ and observation space $\mathsf{Y}$ is a $\mathsf{X}\times\mathsf{Y}$-valued stochastic process $(X_t,Y_t)_{t\geq 1}$ defined on a probability space $(\Omega,\mathcal{A},\mathbb{P})$ such that
\begin{itemize}
\item $(X_t)_{t\geq 1}$ is a Markov chain with state space $\mathsf{X}$.
\item For all $t\geq 1$, the distribution of $Y_t$ given $(X_s)_{s\geq 1}$ only depends on $t$ and $X_t$, and conditional on $(X_s)_{s\geq 1}$, the $(Y_t)_{t\geq 1}$ are independent. Figure \ref{hmm} summarizes this dynamic.
\end{itemize}

\noindent The key point here is that, from a statistical point of view, we do not observe the state sequence $(X_t)_{t\geq 1}$ but only $(Y_t)_{1\leq t\leq n}$. Hence the $X_t$ are called the \emph{hidden states}. The law of the Markov chain $(X_t)_{t\geq 1}$ is determined by its initial distribution $\pi$ such that, for $k\in\mathsf{X}$, $\pi_k=\mathbb{P}(X_1=k)$, and its transition matrices $(Q(t))_{t\geq 1}$ defined, for $i,j\in\mathsf{X}$ and $t\geq 1$, by $Q(t)_{ij} = \mathbb{P}(X_{t+1}=j\mid X_t=i)$. The conditional distributions of $Y_t$ given $X_t$ are called the \emph{emission distribution}. For $t\geq 1$ and $k\in\mathsf{X}$, we shall denote by $\nu_k(t)$ the distribution of $Y_t$ given $X_t=k$. Formally, $Y_t\mid\{X_t=k\}\sim\nu_k(t)$. We assume that for any $t\geq 1$ and $k\in\mathsf{X}$, $\nu_k(t)$ is absolutely continuous with respect to some dominating measure $\mu$ defined on $\mathcal{Y}$ and we denote by $f_{k,t}:\mathsf{Y}\longrightarrow\mathbb{R}_+$ the corresponding density, which will be refered to as the \emph{emission density}. When the transition matrices and emission distributions are constant over time, the corresponding model is sometimes called \emph{homogeneous hidden Markov model}, or just \emph{hidden Markov model} when there is no ambiguity.

\paragraph{Parametric framework}

Assume that the transition matrices and the emission distributions depend on a parameter $\theta\in\Theta$, where $\Theta$ is a compact subset of $\mathbb{R}^q$, for some $q\geq 1$. We now precise the way they depend on $\theta$.
\begin{itemize}
\item The function $t\mapsto Q(t)$ belongs to a known parametric family indexed by an unknown parameter $\beta$ to be estimated.
\item For any $t\geq 1$, the emission distributions $\nu_{k}(t)$ belong to a known (not depending on $t$) parametric family (e.g. gaussian) indexed by a parameter $\theta^Y(t)$. In addition, we assume that the function $t\mapsto\theta^Y(t)$ itself belongs to a known parametric family (e.g. affine functions), with parameter $\delta$.
\item Thus we can write $\theta=(\beta,\delta)$. We will denote by $Q^\theta(t)$ the transition matrix at time $t$ when the parameter is $\theta$ and, for $1\leq k\leq K$,  $f_{k,t}^\theta$ the $k$-th emission density at time $t$ when the parameter is $\theta$.
\end{itemize}

\noindent Note that we consider that the number of states $K$ is known, although this is not the case in practice. We discuss this issue in section \ref{inference}. In this framework we proceed to the estimation of the parameters using maximum likelihood inference. Having observed $(Y_1,\dots,Y_n)$, the likelihood of the model when the parameter is $\theta$ and the initial distribution is $\pi$ is given by 
$$p^{\theta,\pi}(Y_1,\dots,Y_n)=\sum_{x_1,\dots,x_n}\pi_{x_1}Q^\theta(1)_{x_1x_2}f^\theta_{x_1,1}(Y_1)Q^\theta(2)_{x_2x_3}f^\theta_{x_2,2}(Y_2)\dots Q^\theta(n-1)_{x_{n-1}x_n}f^\theta_{x_n,n}(Y_n).$$
Then we define the maximum likelihood estimator (MLE):
$$\hat{\theta}_{\pi,n}=\argmax_{\theta\in\Theta}p^{\theta,\pi}(Y_1,\dots,Y_n).$$
The practical computation of the MLE can be performed using the well-known Expectation Maximization (EM) algorithm. See section \ref{inference} for the details of this algorithm in our framework.

\paragraph{Theoretical guarantees} The statistical properties of hidden Markov models have been studied extensively since the 1960's. However, general identifiability conditions have only been proved recently. Following \cite{allman2009}, the authors of \cite{gassiat2016} proved that stationary non parametric HMM are identifiable from the law of three consecutive observations (up to permutation of the states), provided that the emission distribution are linearly independent and that the transition matrix has full rank. \cite{alexandrovich2016} prove a similar result with slightly weaker assumptions.  The properties of the maximum likelihood estimator in homogeneous HMM are now well-known. Its strong consistency has first been established by \cite{baum1966} in the case where both the state space and the observation space are finite. This result has been generalized to continuous observation spaces by \cite{leroux1992}. See also \cite{douc2011} and references therein. The literature is less abundant when it comes to non-homogeneous hidden Markov models. In \cite{ailliot2015consistency} and \cite{pouzo2016}, the authors consider models where the observation distribution depends not only on the current state, but also on previous observations, and where the transition matrices depend on the previous observations. They prove the consistency of the maximum likelihood estimator in this framework. More recently, \cite{diehn2018} study the case of non-homogeneous hidden Markov models that can be asymptotically approximated by an homogeneous one. They introduce a quasi-maximum likelihood estimator and prove its consistency. None of the previously stated results apply to the model we introduce in this paper. In \cite{touron2018} however, the author obtains the consistency of the maximum likelihood estimator in HMM with periodic transition matrices and emission distributions, which applies to our model in the special case where the temperature trends are zero, or to the precipitation process alone. There is also an on-going work about HMM with trends. 

\section{Application to precipitation and temperature}\label{sec3}

\subsection{Data}

\paragraph{Description of the data}\label{data-description}

\begin{figure}[H]
\centering
\includegraphics[scale=0.4]{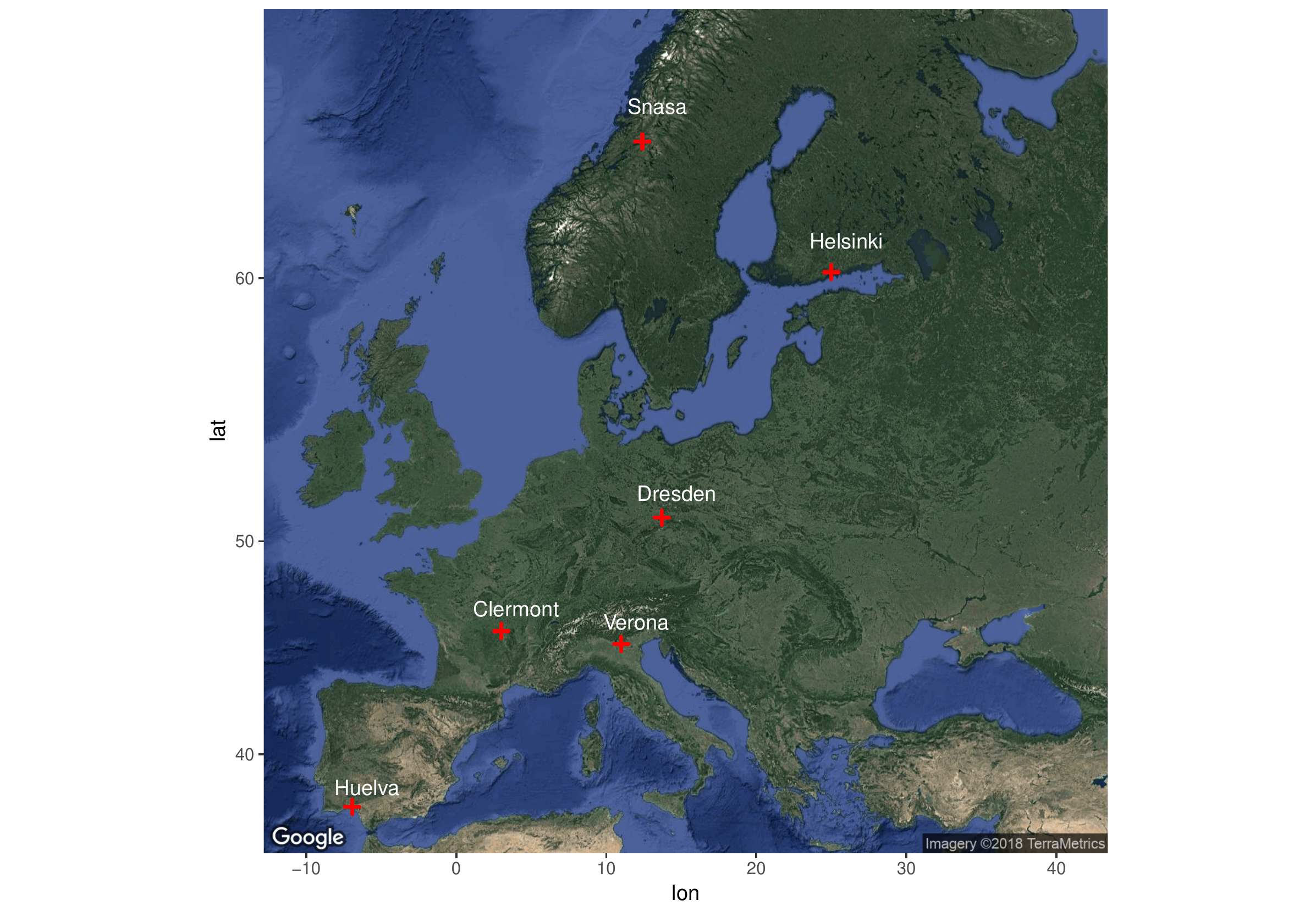}
\caption{Stations locations}\label{map}
\end{figure}

\begin{figure}[H]
\centering
\includegraphics[scale=0.6]{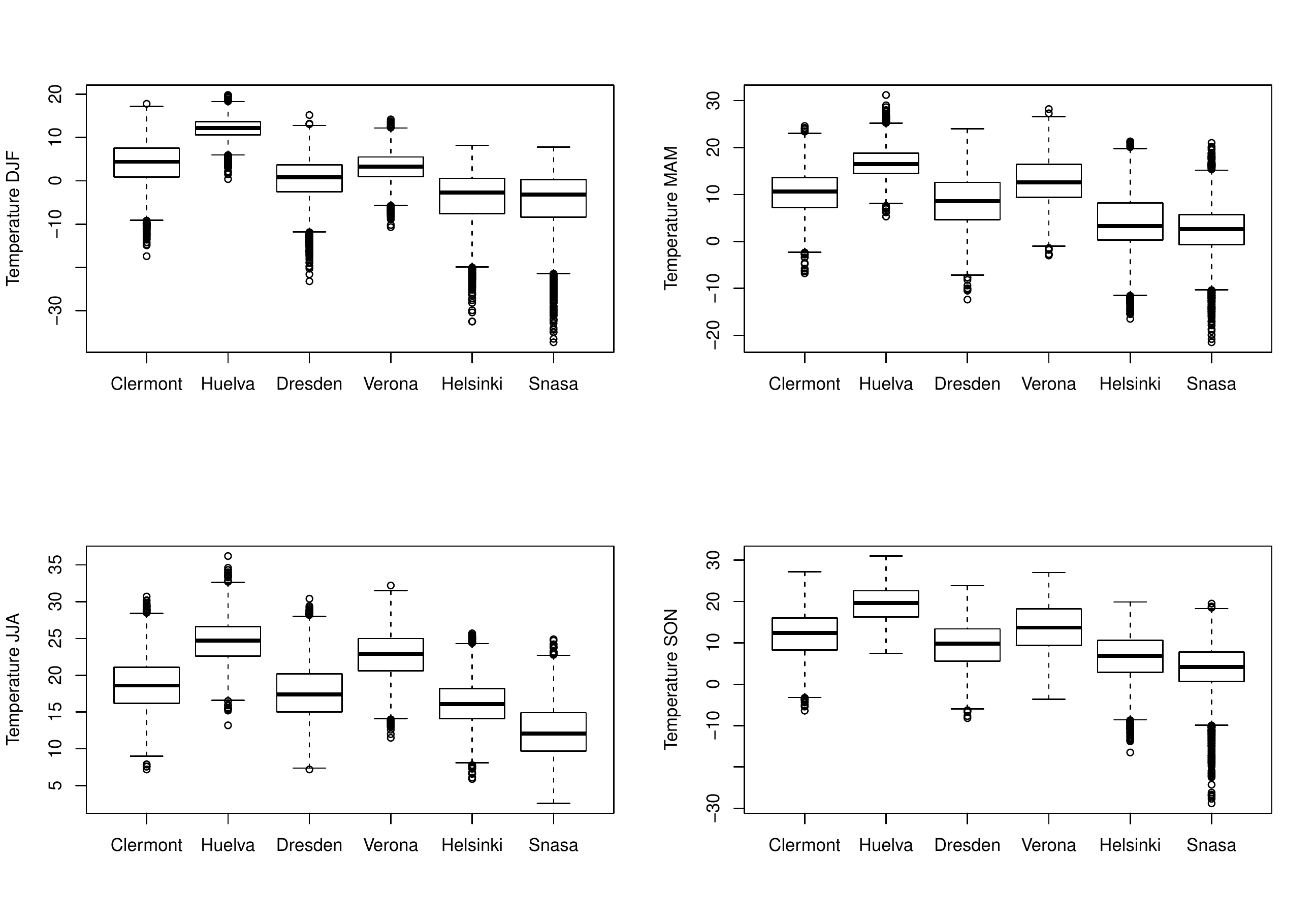}
\caption{Boxplots of the temperature at the different locations for each season}\label{boxplot_temp_saison}
\end{figure}

We used data from the \emph{European Climate Assessment and Datasets} (ECA\&D) project\footnote{Data freely available at \url{https://www.ecad.eu//dailydata/index.php}}. Six weather stations were considered: Helsinki (Finland), Dresden (Germany), Verona (Italy), Huelva (Spain), Clermont-Ferrand (France) and Sn\aa sa (Norway). For each station, the data consists of mean daily temperature and daily precipitation from 1954/01/01 to 2014/12/31. Figure \ref{map} presents the locations of the weather stations, whereas Figure \ref{boxplot_temp_saison} shows that the distribution of temperature in the 6 stations differ in many ways: mean, variance, range, skewness, seasonal behaviour... 

\noindent Table \ref{stat_precip} presents some basic statistics concerning precipitations at the 6 stations under study. Mean yearly precipitations range from 501.5 mm in Huelva (Spain) to 964.2 mm in Sn\aa sa (Norway), where the precipitation frequency is $0.62$ compared to only $0.19$ in Huelva. Also, the maximum observed daily precipitation in Sn\aa sa is 65.9 mm, compared to 198 mm in Verona. The mean value of (non-zero) daily precipitations ranges from 3.4 mm in Helsinki to 7.3 mm in Huelva.

\begin{table}[ht]
\centering
\begin{tabular}{|l|rrrrrr|}
  \hline
 & Clermont & Huelva & Dresden & Verona & Helsinki & Snasa \\ 
  \hline
Mean yearly precipitation (mm) & 584.2 & 501.5 & 665.3 & 803.3 & 638.9 & 964.2 \\ 
  Max. observed precipitation (mm) & 75.3 & 160.0 & 158.0 & 198.0 & 79.3 & 65.9 \\ 
  Precipitation frequency & 0.40 & 0.19 & 0.49 & 0.34 & 0.51 & 0.62 \\ 
  Mean positive precipitation (mm) & 4.0 & 7.3 & 3.7 & 6.5 & 3.4 & 4.2 \\ 
   \hline
\end{tabular}
\caption{Basic statistics for precipitation}\label{stat_precip}
\end{table}

We chose on purpose weather stations where climate strongly differs in order to test the robustness of our model when applied to different climates.

\paragraph{Seasonalities and trends}

\begin{figure}[H]
\centering
\includegraphics[scale=0.5]{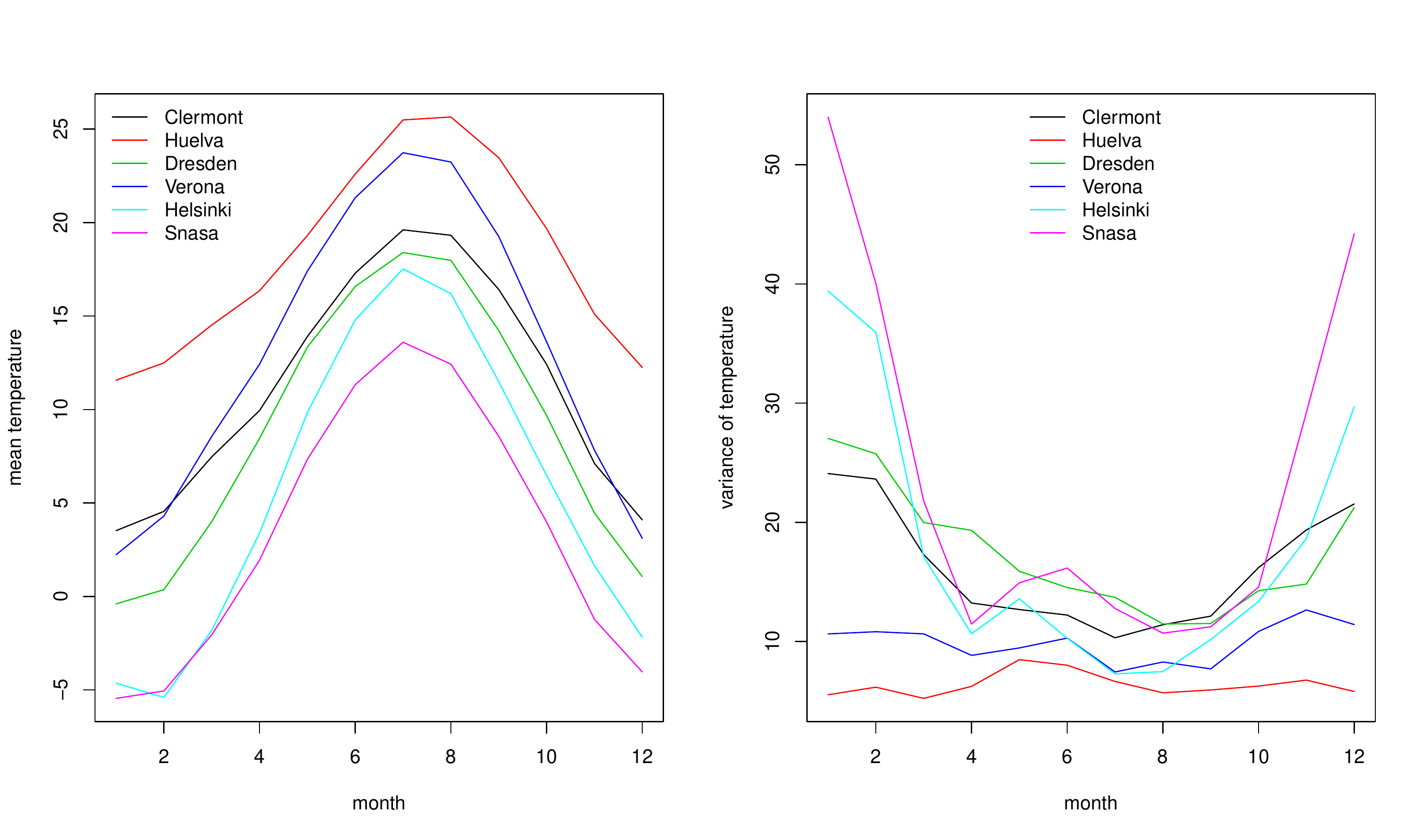}
\caption{Monthly mean and variance of temperature}\label{temp_saison}
\end{figure}

When observed at a daily time step temperature and precipitation times series are not stationary, in that their distribution varies through time. Temperature obviously exhibits a seasonal cycle, as shown on the left panel of Figure \ref{temp_saison}. The right panel of Figure \ref{temp_saison} shows the seasonality of the variance of temperature, which is rather flat in the two stations of Southern Europe (Verona and Huelva). We also notice that these stations have the lowest variances. On the opposite, the variance of temperature displays a very clear seasonality in the stations of Northern Europe (Sn\aa sa and Helsinki), the variability being much higher in the winter months.

\begin{figure}[H]
\centering
\includegraphics[scale=0.6]{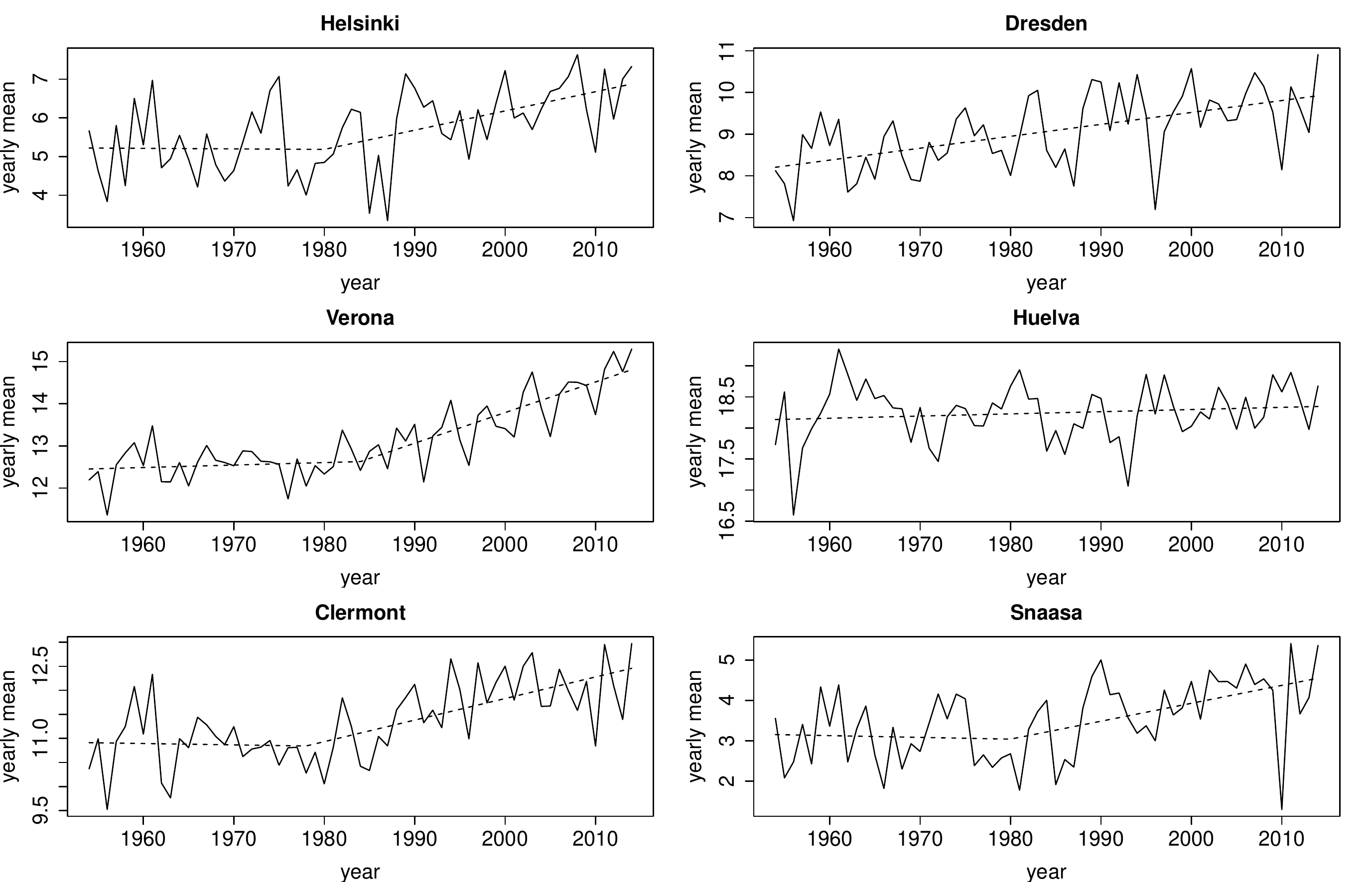}
\caption{Yearly means of temperature (black) and associated optimal trends (blue)}\label{fig-tend-temp}
\end{figure}

The non-stationarity of temperature is also caused by the existence of a trend, corresponding to global warming. The black lines in Figure \ref{fig-tend-temp} are the yearly mean temperature for each site. All sites except Huelva seem to exhibit an increasing trend. However, the shape and the slope of this trend differ among the different sites. For stations like Helsinki or Dresden it seems reasonable to consider a linear trend over the whole observed period, whereas Verona or Clermont exhibit a change point in the warming rate. Thus the modeling of the trend should be site-specific. Figure \ref{fig-tend-temp} suggests that a simple parametric form for the trends could be linear or piecewise linear with two pieces. Then for each site, two questions arise: 
\begin{itemize}
\item is there a breaking point (i.e. a change in the slope of the trend)?
\item if yes, where is it?
\end{itemize}
To answer these questions, we can use the yearly mean temperatures $(\bar{Y}_a)_{a \in \{1954,\dots,2014\}}$. The linear regression is then given by

$$(\hat{\alpha}^{LM},\hat{\beta}^{LM}) = \argmin_{\alpha,\beta\in\R}\sum_{a=1954}^{2014}(\bar{Y}_a - \alpha a - \beta )^2$$
In a piecewise linear model, the optimal breaking point can be found by computing

$$\hat{\tau} = \argmin_{1954\leq \tau \leq 2014} \min_{\alpha,\beta,\gamma\in\R} \sum_{a=1954}^{2014}(\bar{Y}_a - \alpha a - \beta - \gamma(a-\tau)\mathbf{1}_{t>\tau})^2.$$
Then the corresponding piecewise linear regression is given by

$$(\hat{\alpha}^{PLM},\hat{\beta}^{PLM},\hat{\gamma}^{PLM}) = \argmin_{\alpha,\beta,\gamma\in\R}\sum_{a=1954}^{2014}(\bar{Y}_a - \alpha a - \beta - \gamma(a-\hat{\tau})\mathbf{1}_{t>\hat{\tau}})^2.$$
In order to test for the significance of the breaking point, we perform a likelihood ratio test: we compute the test statistic
$$\Lambda = 2\left(\log\mathcal{L}(\hat{\alpha}^{PLM},\hat{\beta}^{PLM},\hat{\gamma}^{PLM}) - \log\mathcal{L}(\hat{\alpha}^{LM},\hat{\beta}^{LM})\right),$$
with $\mathcal{L}$ the likelihood function, where we considered gaussian residuals (this assumption was tested with a Kolmogorov-Smirnov test). Then $\Lambda$ is compared to a quantile of the $\chi^2$ distribution with one degree of freedom.

\begin{table}[H]
\centering
\begin{tabular}{|c|c|c|c|}
\hline 
\textbf{Station} & \textbf{Test result} & \textbf{p-value} & $\tau$ \\ 
\hline
Helsinki & PL & $0.049$ & 1980 \\ 
\hline 
Dresden & L & $0.16$ & - \\
\hline
Verona & PL & $3.10^{-6}$ & 1987\\
\hline
Huelva & PL & $0.026$ & 1961\\
\hline
Clermont & PL & $0.012$ & 1978\\
\hline
Sn\aa sa & PL & $0.047$ & 1980\\
\hline
\end{tabular} 
\caption{Test of the parametric form of the trends}\label{table_test}
\end{table}

Table \ref{table_test} gives the results of this procedure for the six sites. In the "test result" column, PL means that the test rejected the linear model at the risk level $0.05$ (which means that the trend is piecewise linear), and $L$ means that the test did not reject the linear model. The last column $\tau$ is the year of the change in the slope of the trend, when there is such a change. Thus, the test rejects the simple linear trend for all sites but Dresden. For piecewise linear trends, the breaking point is in the decade 1978-1987, which is consistent with climatology, except for the site of Huelva (1961), surprisingly. The optimal trends are depicted in Figure \ref{fig-tend-temp}. We see that as far as Huelva is concerned, despite the result of the test, the piecewise linear trend cannot be considered significant. First, the change point in 1961 is not consistant with climatology, as it would correspond to a warming stopping in 1961. Then, we see that the procedure described above was misleaded by the unusually cold year 1956. For these reasons, we choose a simple linear trend for this station.

\begin{figure}[H]
\centering
\includegraphics[scale=0.5]{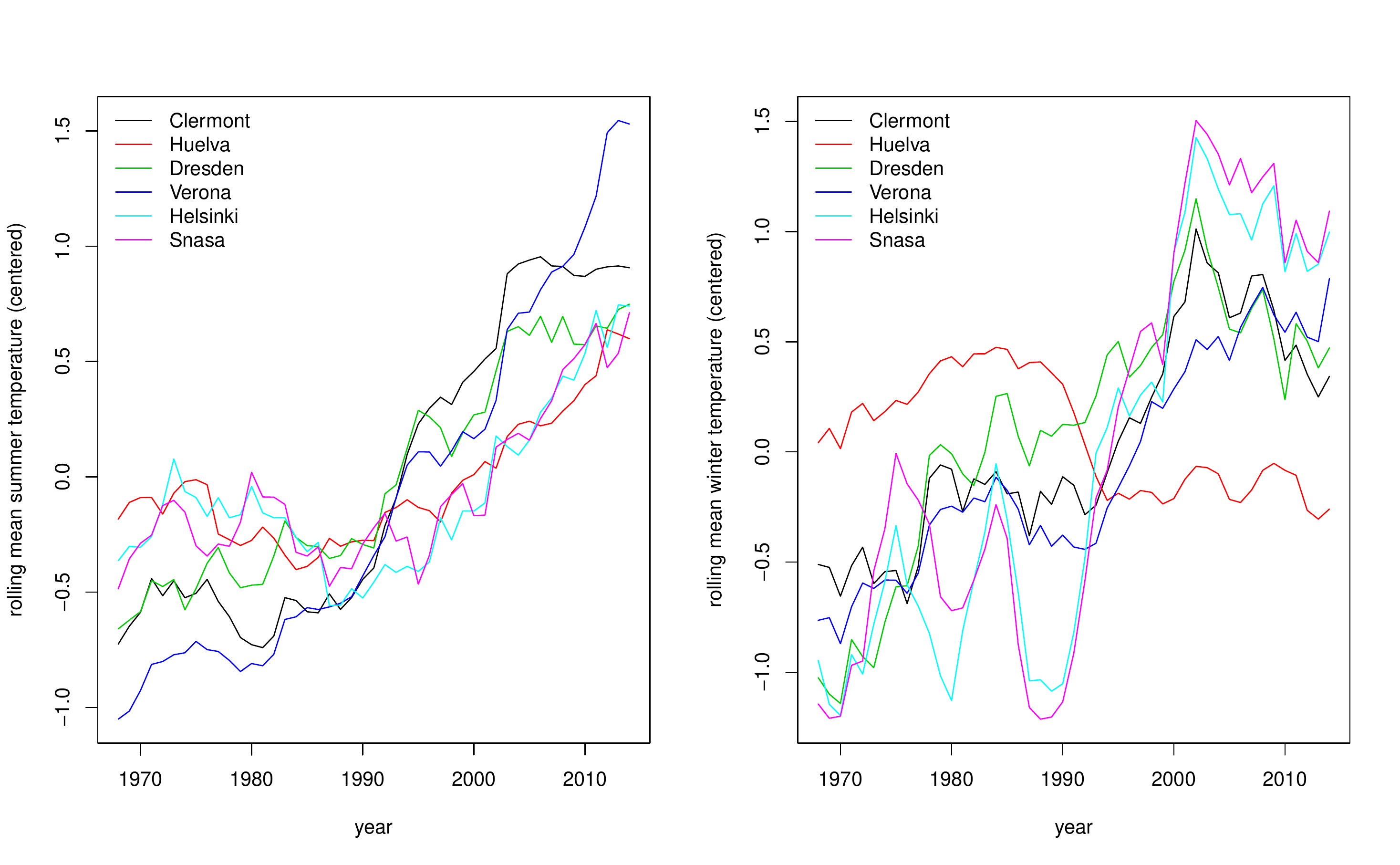}
\caption{Temperature trend in summer (left) and winter (right)}\label{fig-tend2}
\end{figure}

Furthermore, climate change does not affect summer and winter the same way. For each location, we computed the mean temperature of summer months (June, July, August) and we centered it to remove the shift between the stations. We then applied a rolling mean with a window width of 15 years in order to smooth the effect of interannual variability. Thus the value corresponding to the year 1968 is the mean over the period 1954-1968. We performed the same operation considering winter months (December, January, February). The results can be seen in Figure \ref{fig-tend2}. In summer, all the stations show an increasing trend. However, the shape of the trend may differ according to the stations. Also, the amplitude of the warming is higher in Verona ($2.5^\circ$C) than in Helsinki and Sn\aa sa (about $1^\circ$C). In winter, the amplitude of the warming is higher in Northern Europe (Helsinki, Sn\aa sa) than in the South (Verona, Huelva). We actually do not notice any warming in Huelva. Our model will deal with this phenomenon using seasonal transition probabilities between the states, and state-dependent trends (see Section \ref{modele})


\begin{figure}[H]
\centering
\includegraphics[scale=0.5]{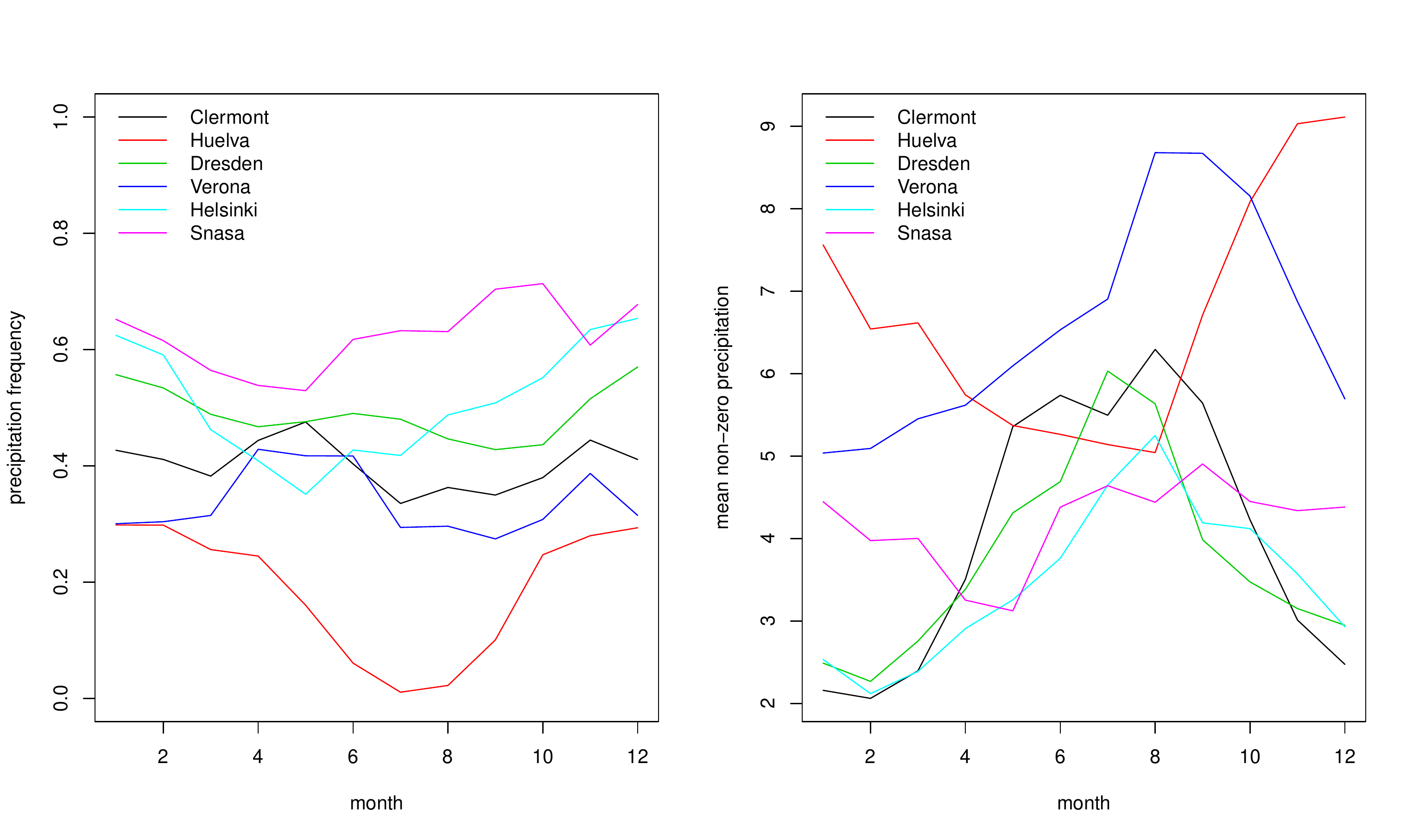}
\caption{Seasonality of precipitation}\label{precip_saison}
\end{figure}

As for precipitation, both the occurrence process (frequency of precipitation) and the intensity process have a seasonal behaviour, as shown in Figure \ref{precip_saison}. The left panel displays the frequency of precipitation, month by month, for each station. Here again, we observe that the shape as well as the amplitude of this seasonality depend on the location. In Helsinki for example, the frequency of precipitation is at its lowest in spring and at its highest in winter. It's the opposite in Verona. The station of Huelva becomes very dry in summer and its maximum precipitation frequency is reached in January (only 0.3). The right panel in Figure \ref{precip_saison} displays the mean value of non-zero precipitation, month by month, for every site. Observe that this quantity exhibits a strong seasonal behaviour (except in Sn\aa sa) that is different from the one of the occurrence process. Once again, this phenomenon requires some modeling effort that we describe in Section \ref{modele}. 

%
%
%

The existence of a trend in the yearly precipitation amounts can be caused either by a trend in the occurrence process (precipitations become more rare or more frequent) or in this intensity process (the mean value of positive precipitations changes). A slight decrease of the frequency of winter precipitations can be observed in Clermont, whereas the increasing trend in winter in Sn\aa sa is mostly the result of an increasing precipitation intensity (not shown). As these trends are light and concern only two stations, we chose not to include them in the model, which does not degrade the results.\\

\begin{figure}[H]
\centering
\includegraphics[scale=0.5]{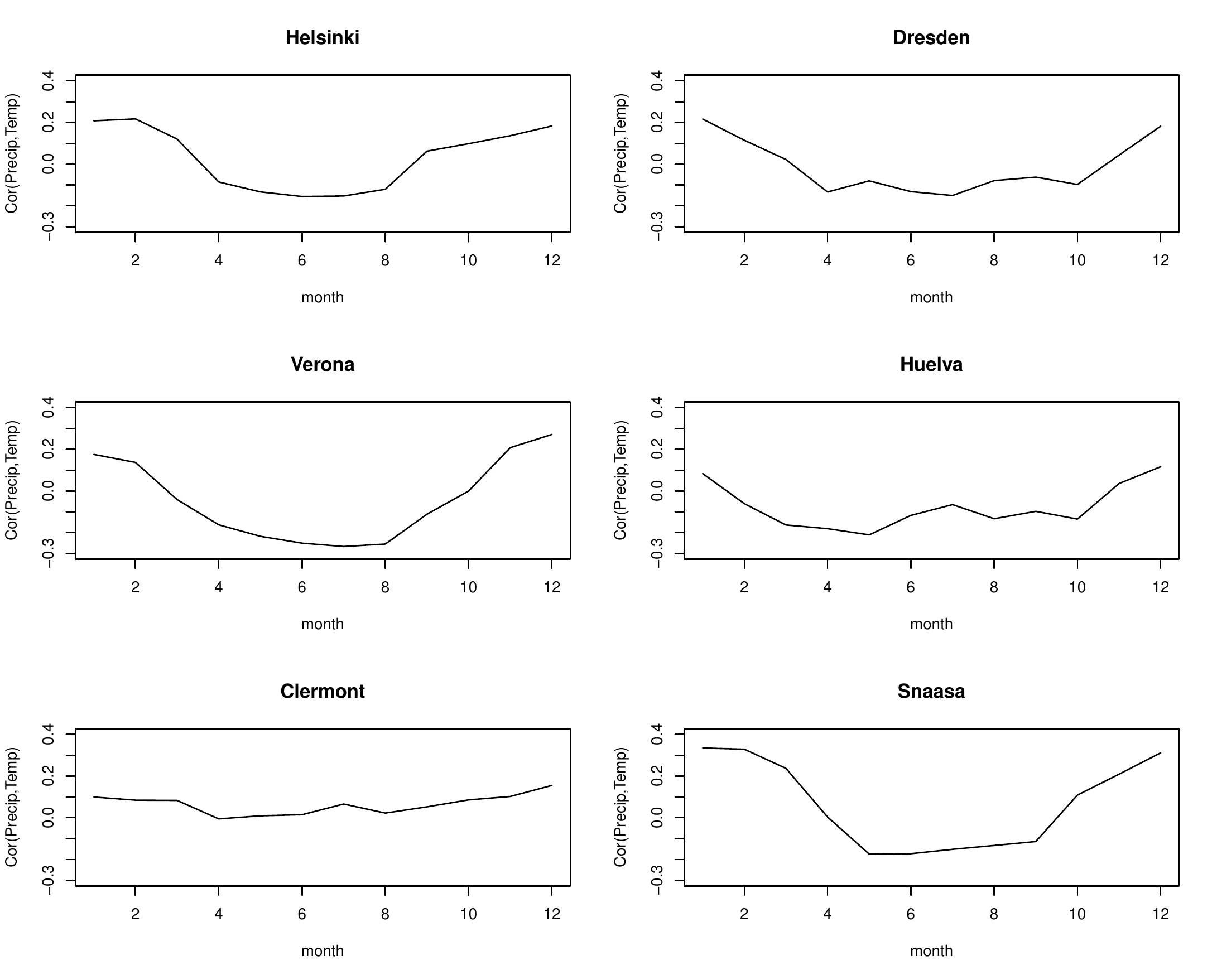}
\caption{Monthly correlations between temperature and precipitations}\label{fig-cor-obs}
\end{figure}

Finally, we highlight the fact that there is also a seasonality in the dependence structure between temperature and precipitation. To see this, we can plot the monthly correlations between these two variables, as in Figure \ref{fig-cor-obs}. All stations but Clermont share a similar shape for this curve, with a negative minimum of correlation in summer, and a positive maximum in winter. This reflects the fact the precipitations mostly occur when the temperature is moderate rather than extreme.

\subsection{Specification of the model}\label{modele}

In section \ref{modgen}, we gave a very general description of our model. In this section, we get more into details as we give the specific forms of the transition structure and the emission distributions. As we want to model simultaneously precipitations and temperature, the observation space is $\mathsf{Y}=\mathbb{R}_+\times\mathbb{R}$ and $Y_t = \left(Y_t^{(1)},Y_t^{(2)}\right)$. The superscript $(1)$ refers to precipitation and $(2)$ refers to temperature.

\paragraph{Transition probabilities}
 The transition matrix at time $t$ in our model is given by

\begin{align}\label{eq-transitions}
Q(t)_{ij} &= \frac{\exp(P_{ij}(t))}{1+\sum_{l=1}^{K-1}\exp(P_{il}(t))}, 1\leq j\leq K-1,\quad 1\leq i\leq K\\
Q(t)_{iK} &= \frac{1}{1+\sum_{l=1}^{K-1}\exp(P_{il}(t))},  1\leq i\leq K.
\end{align}
This is indeed a stochastic matrix, as for all $1\leq i\leq K$, $\sum_{j=1}^KQ(t)_{ij}=1$ and $Q(t)_{ij}>0$. For $1\leq i\leq K$ and $1\leq j\leq K-1$, $P_{ij}$ is a trigonometric polynomial with (known) degree $d$ and period $T$. More precisely,

\begin{equation}\label{eq-transitions2}
P_{ij}(t) = \beta_{ij0}+\sum_{l=1}^d\left(\beta_{ij,(2l-1)}\cos\left(\frac{2\pi}{T}lt\right)+\beta_{ij,2l}\sin\left(\frac{2\pi}{T}lt\right)\right)
\end{equation}

Hence the transition probabilities of the hidden Markov chain are periodic functions of time. Therefore, the relative frequencies of the states will vary through time in a periodic manner. This allows the model to reproduce some of the seasonal behaviours of climate variables.

\paragraph{Emission distributions} In order to allow for flexibility, we choose mixtures as emission distributions. More precisely, the conditional distribution of $Y_t$ given $X_t=k$ is

\begin{equation}\label{loi-emission}
\nu_k(t) = \sum_{m=1}^Mp_{km}\nu^{(1)}_{k,m}(t)\otimes \nu^{(2)}_{k,m}(t)
\end{equation}
The $p_{km}$ are the weights of the mixture, satisfying $p_{km}\geq 0$ and $\sum_{m=1}^Mp_{km}=1$. The component corresponding to precipitations is defined by

\begin{equation}\label{margin-precip}
\nu_{km,t}^{(1)}=\begin{cases}
\delta_0 &, 1\leq m\leq M_1\\
\mathcal{E}\left(\frac{\lambda_{km}}{1+\sigma_k(t)}\right) &, M_1+1\leq m\leq M
\end{cases}, 
\end{equation}
where $\delta_0$ denotes the Dirac mass at $0$, the $\lambda_{km}$ are positive parameters depending on both the state $k$ and the population $m$ in the mixture, and $\sigma_k(\cdot)$ is a state-dependent trigonometric polynomial, modeling the seasonal variations of the intensity of precipitations. Thus the marginal distribution of precipitations in state $k$ at time $t$ is a mixture of a Dirac mass with weight 
$$p_{k0}:=\sum_{m=1}^{M_1}p_{km}$$
and exponential distributions with parameters $\frac{\lambda_{km}}{1+\sigma_k(t)}$ and weights $(p_{km})_{M_1+1\leq m\leq M}$. When $p_{k0}$ is close to $1$, the state $k$ is considered as \emph{dry} whereas it is considered as \emph{wet} when $p_{k0}$ is close to $0$. As the frequency of a given state varies across the year thanks to the seasonal transition probabilities, this will allow the model to capture the seasonal behaviour of the frequency of precipitations. \\

Regarding temperature,

\begin{equation}\label{margin-temp}
\nu_{km,t}^{(2)} = \mathcal{N}\left(T_k(t)+S_k(t)+\mu_{km},\sigma^2_{km}\right),
\end{equation}

so that the marginal distribution of temperature in state $k$ at time $t$ is a mixture of gaussian distributions, with variances $\sigma^2_{km}$.
\begin{itemize}
\item $\mu_{km}$ is a mean parameter that depends on both the state $k$ and the component $m$.
\item $T_k(\cdot)$ is a function corresponding to the temperature trend in state $k$. As shown in Section \ref{data-description}, the parametric form of the trends depends on the sites. We follow the conclusions of Section \ref{data-description} and we choose linear trends for Huelva and Dresden, and piecewise linear trends (with site-specific change points) for the other stations.
\item $S_k(\cdot)$ is a trigonometric polynomial with degree $d$ corresponding to the seasonal cycle of temperature in state $k$.
\end{itemize}

\noindent Note that we allow both the trend and the seasonality of temperature to depend on the hidden state, hence the subscript $k$. Equation \eqref{loi-emission} shows that in each state and each component of the mixture, precipitations and temperature are independent, but of course they are not globally independent. The choices of $K$, $M$, $M_1$ and $d$ are discussed in the next section.

\subsection{Inference of the parameters}\label{inference}

\paragraph{The EM algorithm}

The computation of the maximum likelihood estimator is done using the EM algorithm \citep{dempster77}, which is a generic approach to perform maximum likelihood inference in latent variables models. The details of the algorithm in our particular framework can be found in \cite{touron2018}. Recall that the EM algorithm does not guarantee to find the global maximum of the likelihood function but only a local maximum, depending on its initial point. To overcome this drawback, we launch the algorithm multiple times, using randomly chosen initial points. See also \cite{biernacki2003} where the authors compare several initialization procedures for the EM algorithm.

\paragraph{Model selection}

Our model requires to specify several hyper-parameters:
\begin{itemize}
\item $K$ the number of hidden states,
\item $d$ the degree of the trigonometric polynomials, which sets the complexity of the seasonality,
\item $M$ and $M_1$ which correspond to the complexity of the emission distributions.
\end{itemize}

As the dimensionality of the parameter $\theta$ is a quadratic function of $K$ and a linear function of both $d$ and $M$, the larger these hyper-parameters, the more complex the model is and the better we capture the statistical properties of the data. However, we cannot use too large hyper-parameters, for the following reasons:
\begin{itemize}
\item We may overfit the data.
\item The M step of the EM algorithm requires to solve an optimization problem. As its solution admits no closed form, this is done using a numerical optimization algorithm, which can be difficult and time-consuming if the number of parameters is too large.
\item The likelihood function of models with a large number of parameters may have many sub-optimal local maxima.
\item A large number of hidden states leads to a loss of interpretability of the states, which may be a problem for some practitionners.
\end{itemize}
Therefore, the complexity of the model, especially the number of hidden states, must be chosen carefully. To do so, a standard approach is to use information criteria such as Akaike Information Criterion (AIC), Integrated Completed Likelihood (ICL, see \cite{biernacki2000}) or Bayesian Information Criterion (BIC, see \cite{schwarz1978}). The latter is very popular in applications of HMM, although not justified in theory. The idea of AIC and BIC is to penalize the models with a large number of parameters, in order to realize a trade-off between goodness-of-fit and complexity. If we have to choose a model among a collection $\mathcal{M}$, we minimize over $m\in\mathcal{M}$ the criterion $-\mathcal{L}(\hat{\theta}_m)+\mathrm{pen}(m)$, where $\mathcal{L}(\hat{\theta}_m)$ is the maximum likelihood in the model $m$, and $\mathrm{pen}(m)$ is the penalty associated to the model $m$. For example, in the case of BIC, $\mathrm{pen}(m)=-\frac{p(m)}{2}\log(n)$ where $p(m)$ is the number of parameters of the model $m$ and $n$ is the number of observations. Another approach is cross-validated likelihood (Celeux and Durand, 2008), even though it
is computationally intensive. In \cite{lehericy2016order}, the author introduces a penalized least square estimator
for the order of a nonparametric HMM and proves its consistency. However, when dealing with real world data, other considerations should be taken into account, such as interpretability of the states, computing time, or the ability of the model to reproduce some behaviour of the data, as explained in \cite{bellone2000}. Indeed, according to \cite{pohle2017}, the popular AIC (Akaike Information Criterion) and BIC, as well as other penalized criteria, tend to overestimate the number of states as soon as the data generating process differs from a HMM (e.g. the presence of a conditional dependence), which is often the case in practice. Hence it is advised to use such a criterion as a guide, without following it blindly. Keeping in mind these considerations, we chose to use the BIC criterion to select the number of states, which is by far the most important hyper-parameter because the number of parameters of the model is quadratic in $K$. We found $K=6$ or $K=7$ (depending on the stations) to be good choices. We used our previous experience on univariate models to select $d=2$, $M=4$ and $M_1=2$.

\subsection{Results}

\subsubsection{Estimated parameters}

As we use a hidden Markov model, we do not need to define the states \emph{a priori}, thus we do not need to give them an interpretation before estimating the parameters (e.g. wet or dry state). The determination of the states is data driven and this is one of the perks of HMM. Besides, as our purpose is to produce realistic time series of weather variables, we only need to investigate on the simulations produced by the model, not its parameters. However, it is interesting to take a look at the estimated parameters themselves, thus interpreting the states \emph{a posteriori}. Indeed, interpretability of the states gives credit to the model as it provides a first indication of whether or not it manages to capture some specific meteorological behaviours. We will not provide all the estimated parameters for all of the six stations. Rather, we will give some examples to shed light on the way the different parameters of our model can be physically interpreted. More precisely, the following elements are interesting to look at:

\begin{itemize}

\item Transitions (see Equations (\ref{eq-transitions}) to \eqref{eq-transitions2}):
	\begin{itemize}
	\item transition matrices, specifically the functions $t\mapsto \hat{Q}_{ij}(t)$ for each pair $(i,j)\in \{1,\dots,K\}^2$ and $1\leq t\leq T$. The estimated transition probabilities for the station of Verona are depicted in Figure \ref{fig-transitions}. This provides information on the stability of the states. For example, State $1$ is rather stable in summer, (the probability of remaining in that state is approximately $0.8$), whereas it is much less stable in winter. On the opposite, State $6$ is unstable in summer and more stable in winter.
	\item the relative frequencies of the states, i.e. the probabilities $\mathbb{P}^{\hat{\theta}}(X_t = k)$ for $1\leq k\leq K$, $1\leq t\leq T$. These quantities can be directly obtained through the transition matrices. We represented them for the station of Verona in Figure \ref{fig-loista}. We see that they vary throughout the year, the most frequent states in summer being $1$ and $4$, whereas it is State $3$ in winter.
	\end{itemize}

\begin{figure}[H]
\centering
\includegraphics[scale=0.6]{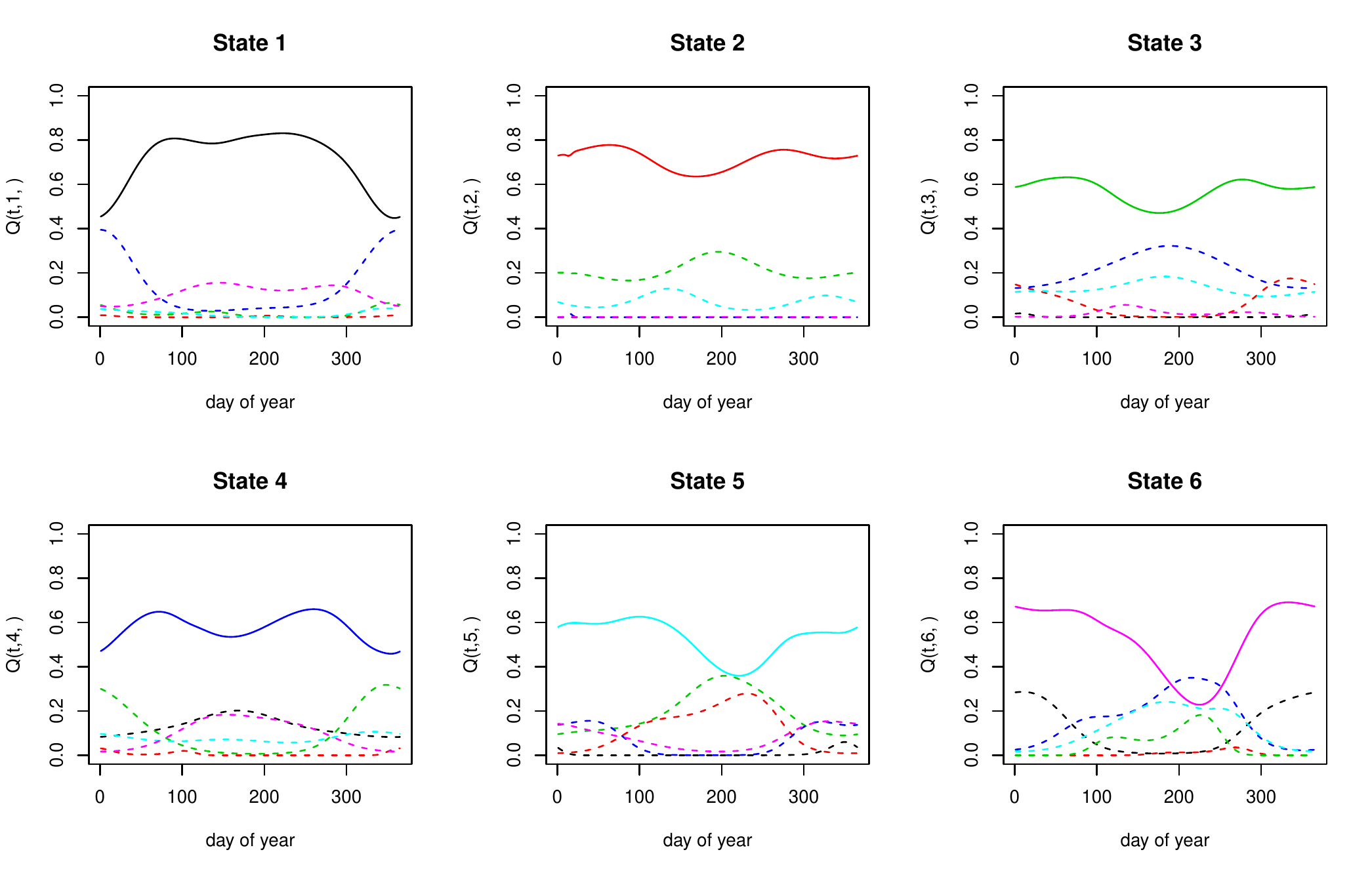}
\caption{Estimated transition probabilities, Verona. Each panel corresponds to a row of the transition matrices. The solid lines represent the diagonal coefficients, i.e. the functions $\hat{Q}_{ii}(\cdot)$ whereas the dashed lines represent the other coefficients. Each color corresponds to a different state.}\label{fig-transitions}
\end{figure}

\begin{figure}[H]
\centering
\includegraphics[scale=0.4]{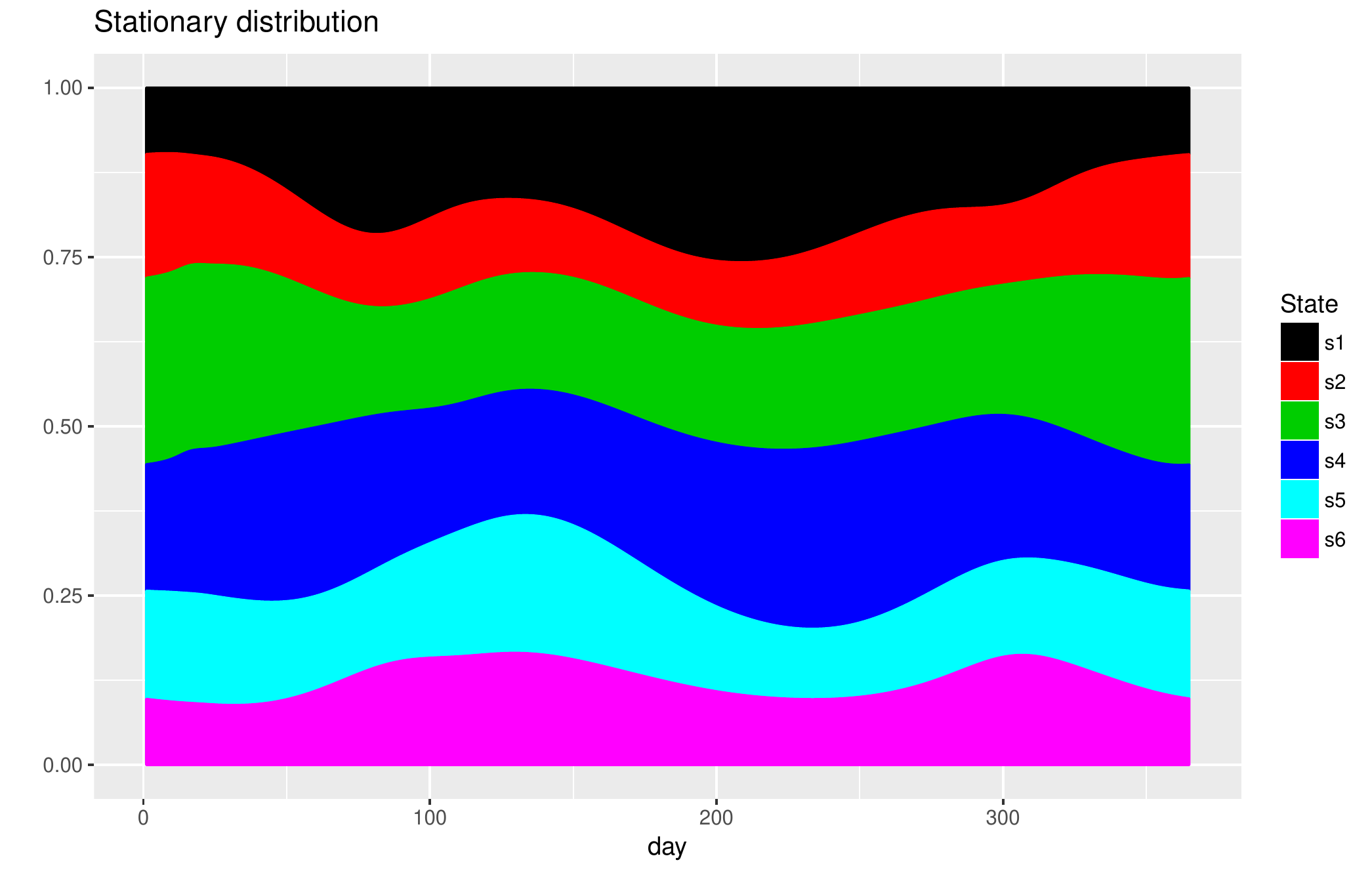}
\caption{Estimated relative frequencies of the states, Verona.}\label{fig-loista}
\end{figure}

\item Temperature parameters (see Equation \eqref{margin-temp}):
	\begin{itemize}
	\item the trends $\hat{T}_k(\cdot)$ for $1\leq k\leq K$. Recall that the trends are modeled as linear for the sites of Huelva and Dresden, and piecewise linear for the others (with site-dependent breaking points) and that we estimate a different trend for each state. The results for the six stations are presented in Figure \ref{fig-trends} and reflect an increase in mean temperature ranging from $0.6^\circ$C in Huelva to $\simeq 2^\circ$C (depending on the state) in Verona.
	\item the state-dependent seasonalities $\hat{S}_k(\cdot)$, corresponding to the yearly cycle of temperature (see Figure \ref{fig-saison}).
	\item the random noise, i.e. the centered gaussian mixture that remains, in each state, when we removed the state-dependent trend and the state-dependent seasonal component. Recall that we chose $M=4$, so that there are $4$ components in each gaussian mixture. We can see in Figure \ref{fig-noise} some peaks in the probability density functions (especially in Dresden). These are caused by estimators $\hat{\sigma}_{km}^2$ that are close to zero. They must be understood as numerical issues in the estimation process (EM algorithm) and have no physical intepretation. Furthermore, they have no negative impact on the quality of the simulated process, as the associated weights $\hat{p}_{km}$ are also close to zero. Figure \ref{fig-noise} also shows that some states have heavier tails (e.g. the state corresponding to the black line in Helsinki). These states can be interpreted as \emph{extreme values states}. Indeed, they induce a larger probability for large deviations from the mean temperature. 
	\end{itemize}

\begin{figure}[H]
\centering
\includegraphics[scale=0.5]{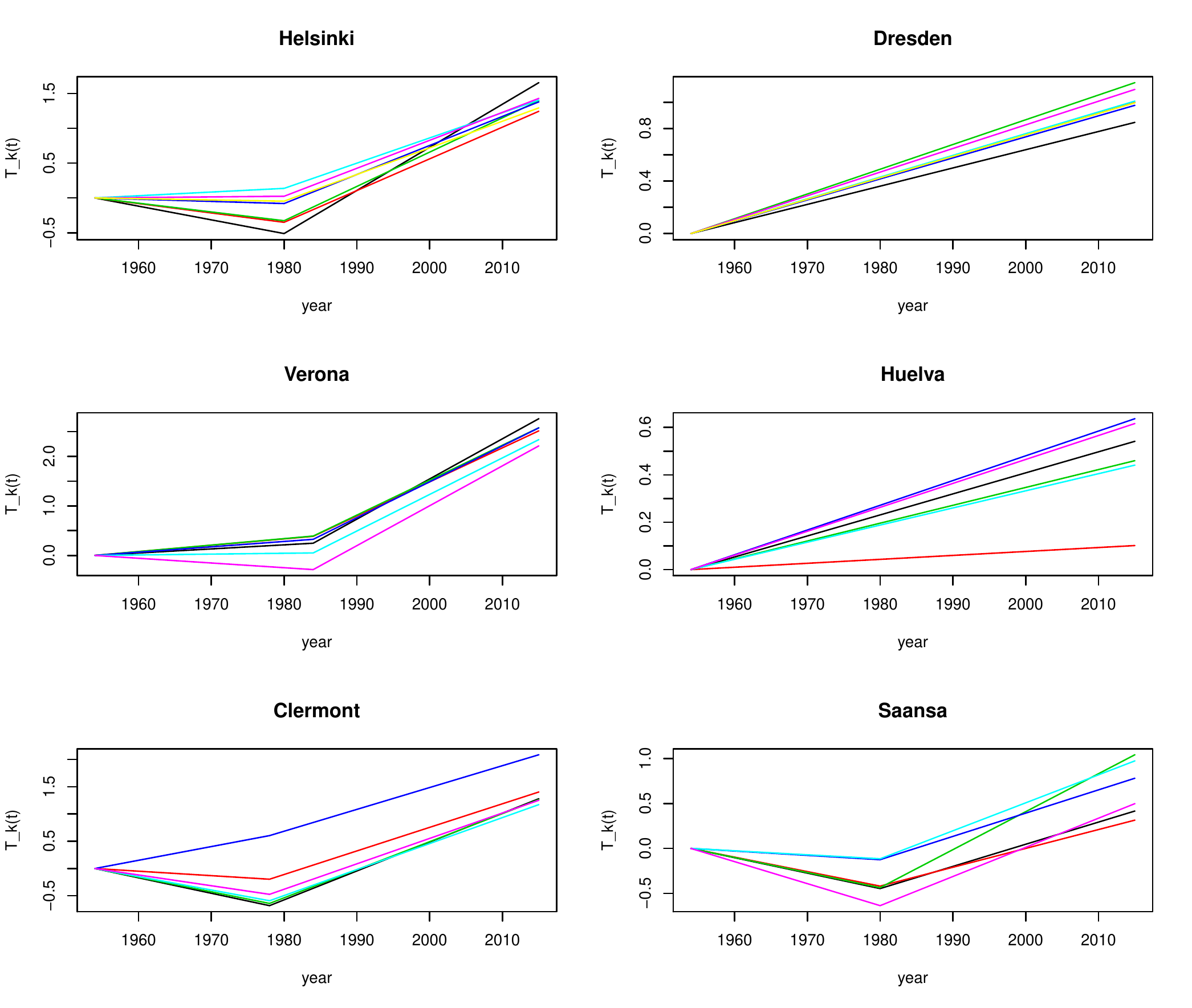}
\caption{Estimated trends (one color per state)}\label{fig-trends}
\end{figure}

\begin{figure}[H]
\centering
\includegraphics[scale=0.5]{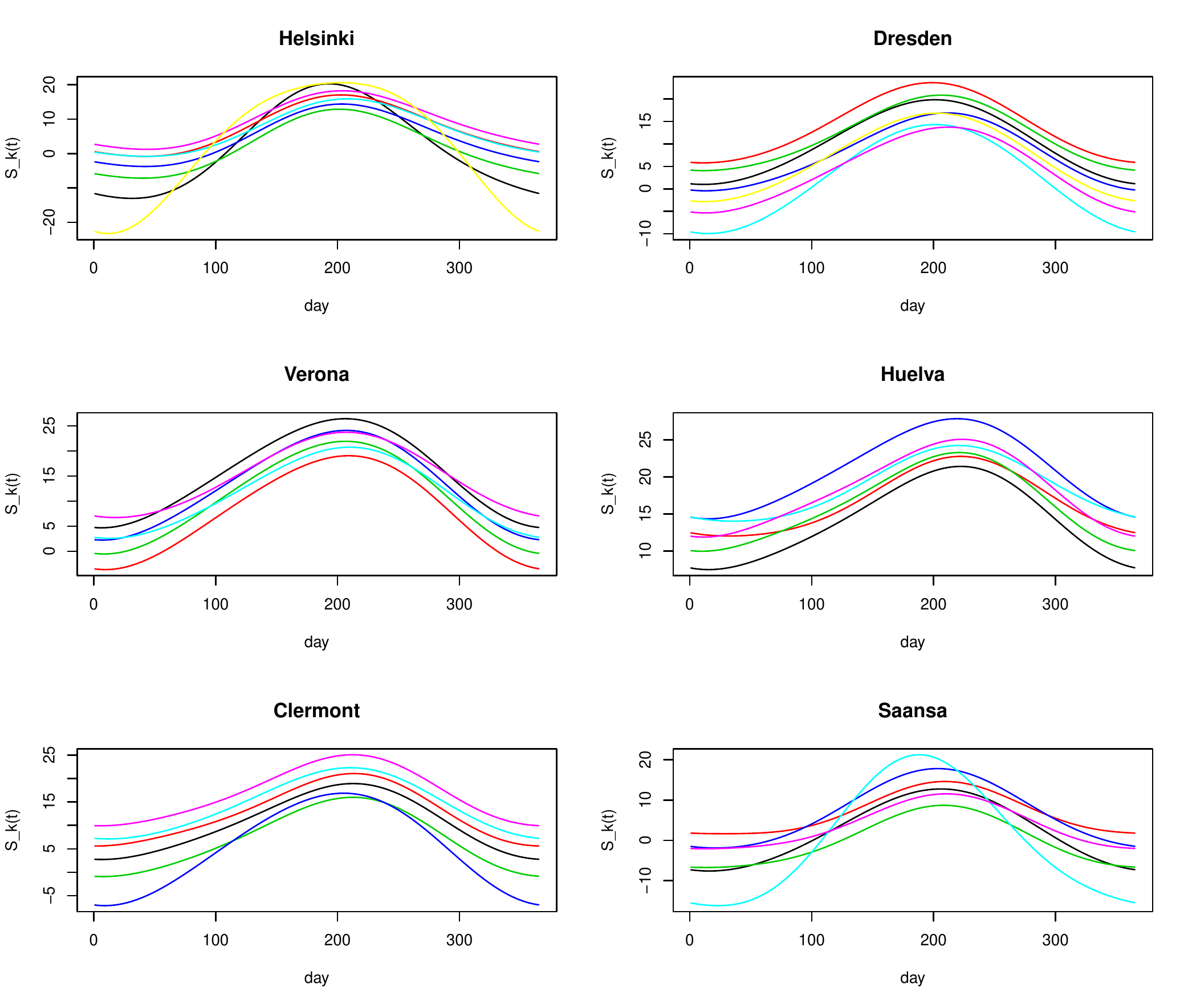}
\caption{Estimated seasonalities of temperature (one color per state)}\label{fig-saison}
\end{figure}

\begin{figure}[H]
\centering
\includegraphics[scale=0.5]{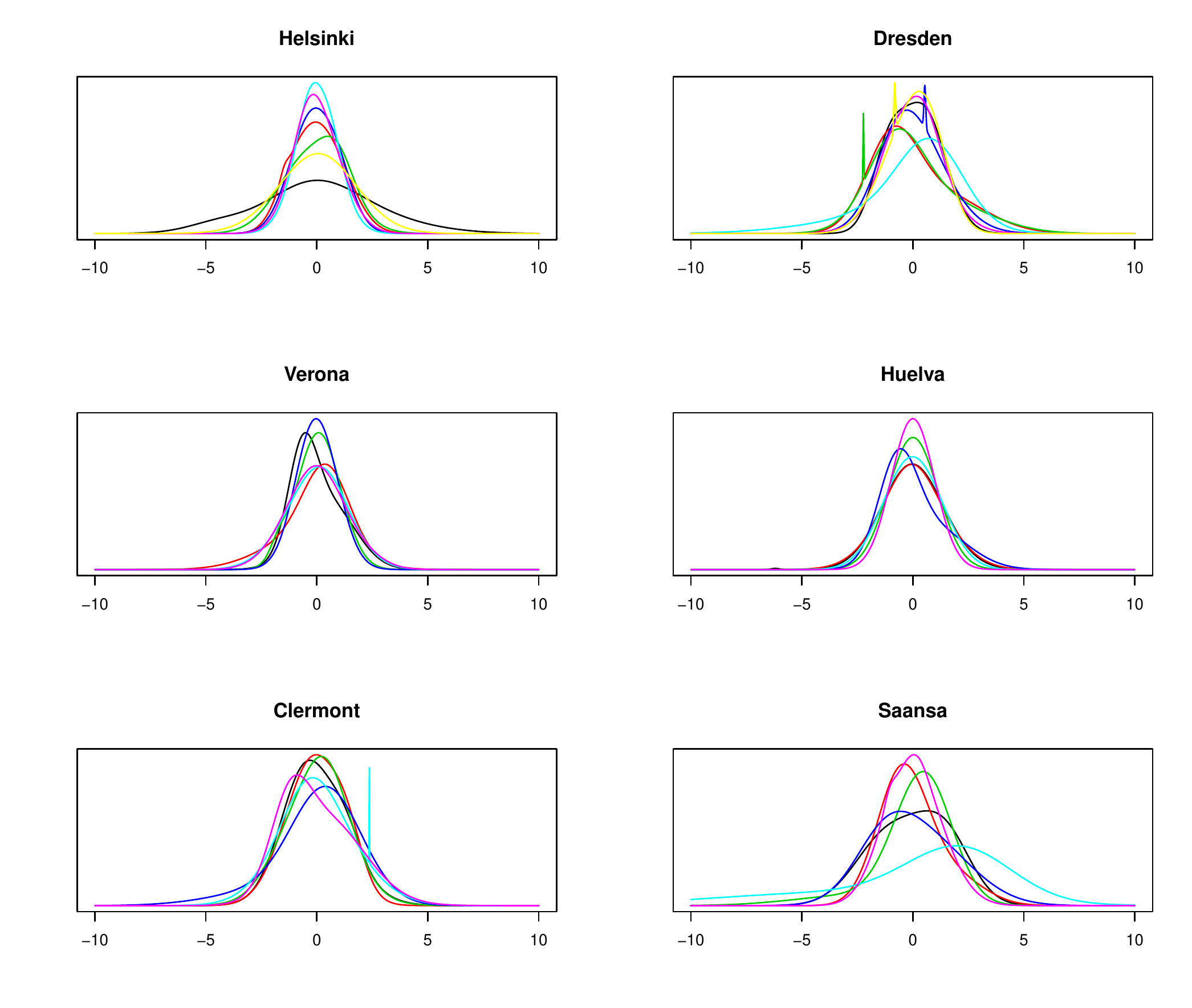}
\caption{Probability density functions of the random noises of the temperature (one color per state)}\label{fig-noise}
\end{figure}

\item Precipitation parameters (see Equation \eqref{margin-precip}):
	\begin{itemize}
	\item the estimated seasonal components $1+\hat{\sigma}_k(t)$: the larger they are, the heavier the precipitations, if any. They are depicted in Figure \ref{fig-saisonsP}. One can see that for all the considered sites, some of the states clearly exhibit a seasonal behaviour regarding the intensity of precipitations, in accordance with the right panel of Figure \ref{precip_saison}. 
	\item the weights of the Dirac masses, i.e. $\left(\hat{p}_{k0}\right)_{1\leq k\leq K}=\left(\sum_{m=1}^{M_1}\hat{p}_{km}\right)_{1\leq k\leq K}$: how dry or wet are the states.  This can be compared to the left panel of Figure \ref{precip_saison}. For example, for the station of Verona,
	$$\left(\hat{p}_{k0}\right)_{1\leq k\leq K} = (0.88, 0.68, 0.93, 0.95, 0.09, 0.09),$$
	so that the states $1$, $3$ and $4$ are mostly dry, whereas the states $5$ and $6$ are rainy.
	\item the estimated parameters of the exponential distributions involved in each state, i.e. 
	$$\left(\hat{\lambda}_{km}\right)_{\substack {1\leq k\leq K\\ M_1+1\leq m\leq M}}.$$
	 They represent the baseline intensity of rainfall (independently from the seasonal variations).
	\end{itemize}

\begin{figure}[H]
\centering
\includegraphics[scale=0.5]{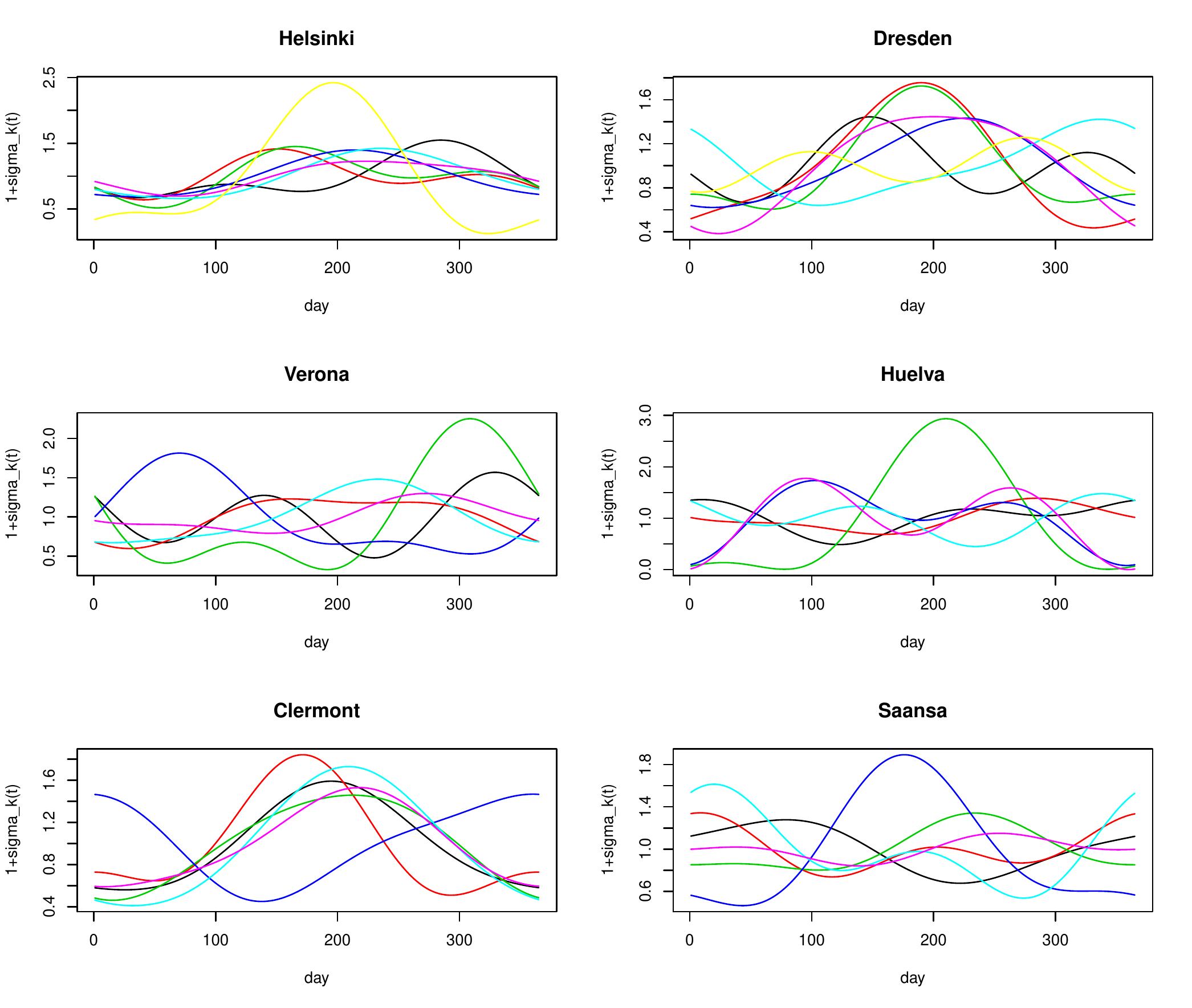}
\caption{Seasonalities in the precipitation intensity (one color per state)}\label{fig-saisonsP}
\end{figure}

\end{itemize}

Until now, we have not tested the model using simulations, but the estimated parameters are consistent with what we would expect from precipitations and temperatures. The validation of the model through simulations is the purpose of the next paragraph.

\subsubsection{Validation of the model}



\paragraph{Simulation} At this stage, for each site, we have ran the EM algorithm using our precipitation and temperature data as inputs, and we have at our disposal a vector of estimated parameters $\hat{\theta}$. This vector of parameters can be used to produce synthetic time series of temperature and precipitations $\left(Y_t^{\mathrm{sim}}\right)_{1\leq t\leq n} = \left(Y_t^{(1),\mathrm{sim}},Y_t^{(2),\mathrm{sim}}\right)_{1\leq t\leq n}$, where $n$ is the length of our observed time series. The simulation procedure is the following. 

\begin{enumerate}
\item Simulate a sequence of states $\left(X_t^{\mathrm{sim}}\right)_{1\leq t\leq n}$ using the estimated transition matrices $\left(\hat{Q}(t)\right)_{1\leq t\leq n}$. The initial state $X_1^{\mathrm{sim}}$ can be chosen arbitrarily or drawn according to the stationary distribution of $\hat{Q}(1)$.
\item Given $\left(X_t^{\mathrm{sim}}\right)_{1\leq t\leq n}$, simulate $\left(Y_t^{\mathrm{sim}}\right)_{1\leq t\leq n}$. At time $t$, if $X_t^{\mathrm{sim}}=k$,
\begin{enumerate}
\item Choose a component $m\in\{1,\dots,M\}$ according to the probability vector $\left(\hat{p}_{km}\right)_{1\leq m\leq M}$.
\item Take $Y_t^{(2),\mathrm{sim}}$ as a realization of a $\mathcal{N}\left(\hat{T}_k(t)+\hat{S}_k(t)+\hat{\mu}_{km},\hat{\sigma}_{km}^2\right)$ distribution. This is our simulated temperature at time $t$.
\item If $1\leq m\leq M_1$, then $Y_t^{(1),\mathrm{sim}}=0$, else take $Y_t^{(1),\mathrm{sim}}$ as a realization of a $\mathcal{E}\left(\frac{\hat{\lambda}_{km}}{1+\hat{\sigma}_k(t)}\right)$. This is our simulated precipitation at time $t$.
\end{enumerate}
\end{enumerate}

Note that the simulation algorithm gives insight into the reason why precipitation and temperature are dependent in our model: they always share the same state $k$, and the same mixture component $m$. This is crucial because we obviously do not want precipiation and temperature to be simulated independently.\\

Thus, using the above algorithm, we are able to simulate very easily and quickly a large number of synthetic time series of the same length as the observed ones. Using a standard laptop, we simulated $N_{\mathrm{sim}}=1000$ independent trajectories of length $n=22265$ in a few minutes, for each site. Now our goal is to compare these simulations to the observed times series, in order to check if our simulations are realistic, with regard to several criteria. Our validation procedure is divided into three parts, aiming to answer the three following questions:
\begin{itemize}
\item is the distribution of the precipitation process $(Y_t^{(1)})_{1\leq t\leq n}$ well reproduced?
\item is the distribution of the temperature process $(Y_t^{(2)})_{1\leq t\leq n}$ well reproduced?
\item is the dependence structure between precipitation and temperature well reproduced?
\end{itemize}

To this aim, we shall consider several statistics, or criteria of validation. These statistics will be computed from the data and from each of the $N_\mathrm{sim}$ simulated trajectory, thus providing a Monte-Carlo estimate of the distribution of each statistic under the model.

\paragraph{Temperature}

\begin{figure}[H]
\centering
\includegraphics[scale=0.5]{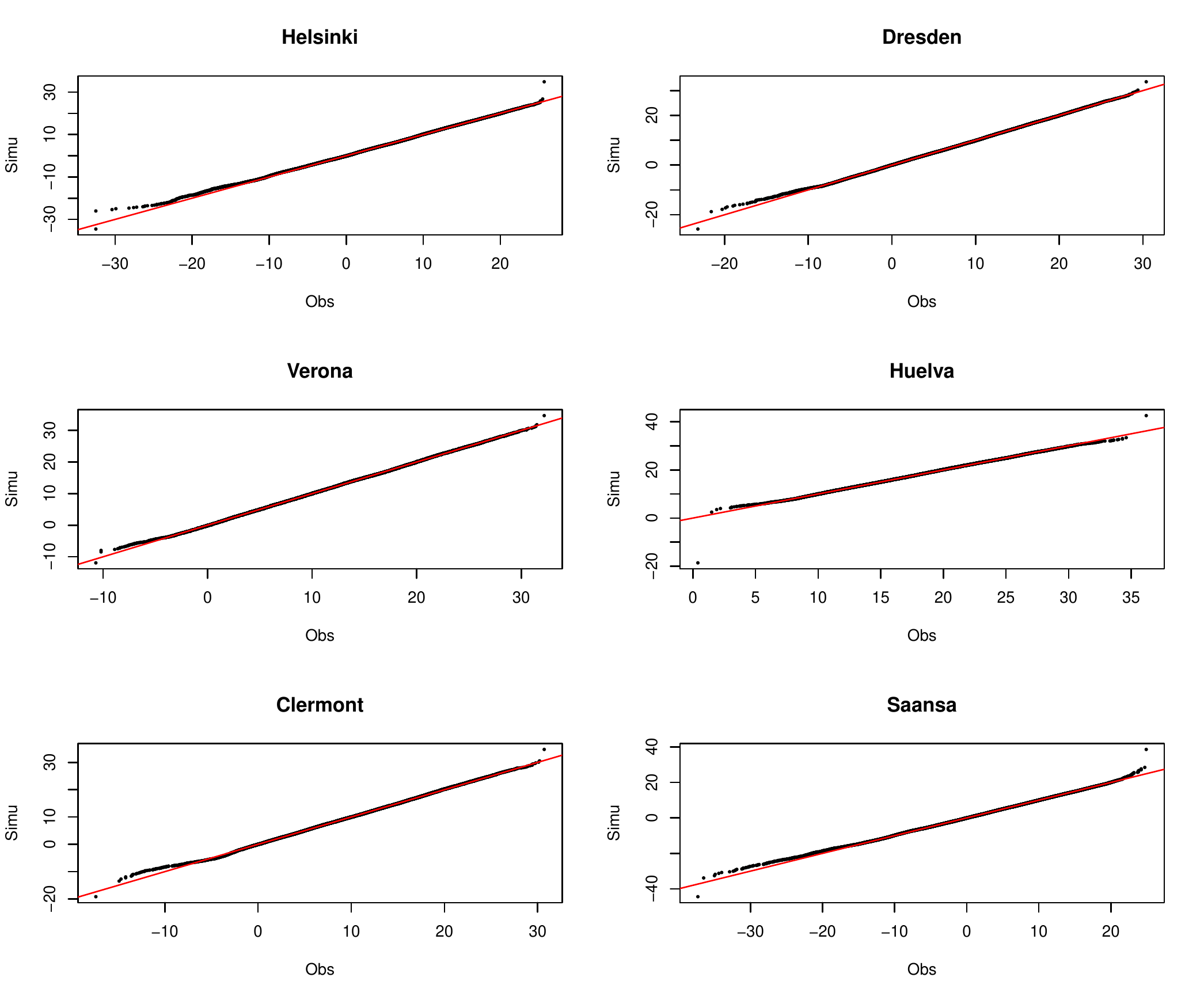}
\caption{Quantile-quantile plots of observed vs simulated temperatures}\label{fig-qqplot-temp}
\end{figure}

Forgetting about the temporal aspect of the temperature time series, we can start by looking at the overall distribution of the temperatures. Figure \ref{fig-qqplot-temp} shows the quantile-quantile plot (QQ-plot) of observed versus simulated temperatures, for each site. Here we mix all the simulated values from the $1000$ trajectories of length $n = 61\times 365 = 22265$. We see that the model can generate values that are more extreme than those observed in the data, both in the left and the right tail of the distribution. In the station of Huelva, the model has generated unrealistically cold values: $61$ values are below $-5$ whereas the minimum observed value is $0.4$. However, these are rare, considering the large number $N_{\mathrm{sim}}\times n$ of simulated values. Apart from that, the overall distribution is well reproduced.

\begin{figure}[H]
\centering
\includegraphics[scale=0.5]{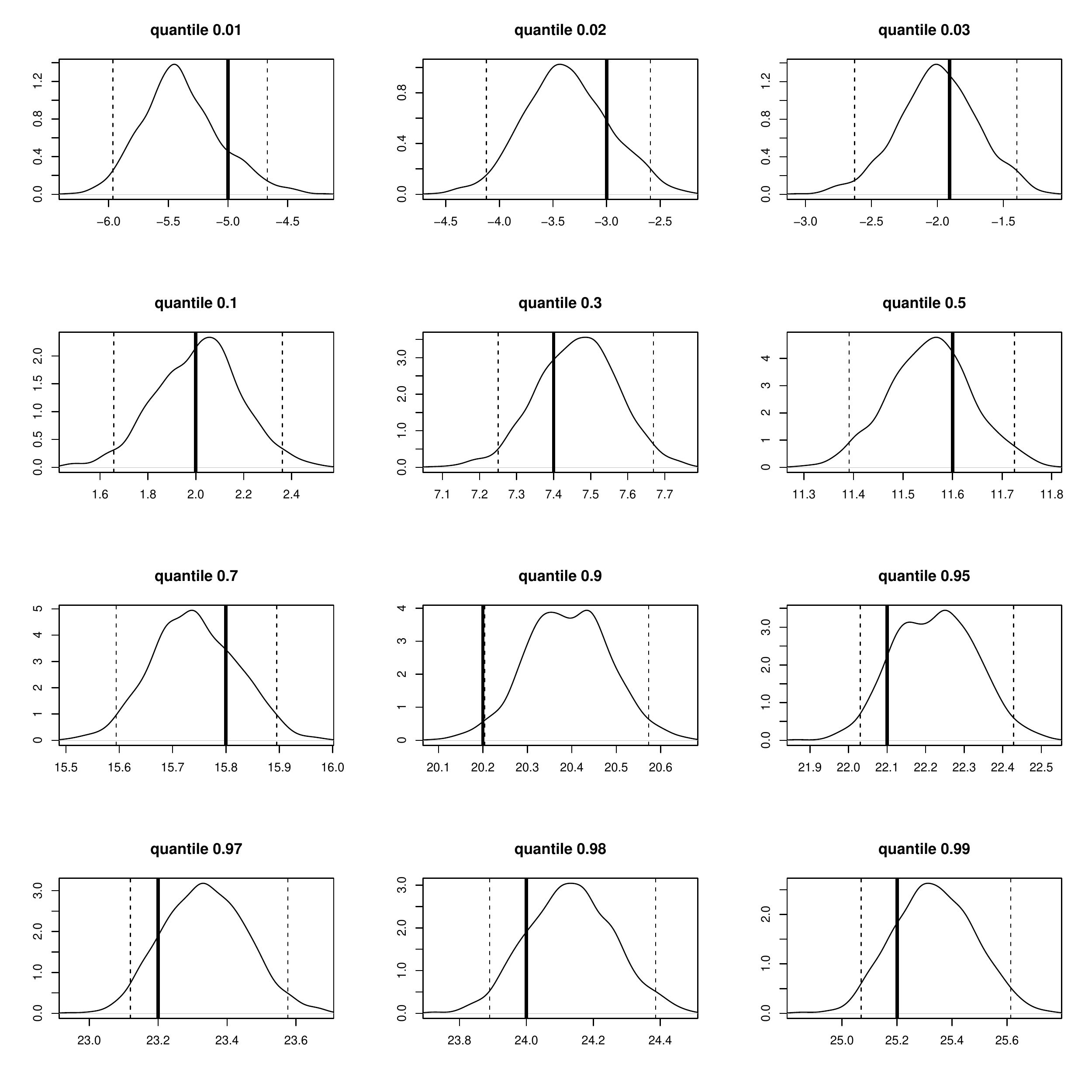}
\caption{Observed (vertical lines) superimposed to the distribution of simulated (curve) quantiles of temperature for the station of Clermont. The dashed vertical lines are the bounds of a $95\%$ confidence interval.}\label{fig-quantiles-temp}
\end{figure}

Another way to look at the overall distribution of temperatures is to compute its observed quantiles and to compare them with the distributions of the quantiles generated by the model. For $\alpha\in (0,1)$, we can compute the observed empirical $\alpha$-th quantile, denoted by $q_\alpha^{\mathrm{obs}}$. Then we perform the same calculation for each simulation $s\in\{1,\dots,N_{\mathrm{sim}}\}$ and we obtain the quantiles $q_\alpha^s$. Finally, $q_\alpha^\mathrm{obs}$ is compared to the distribution of $(q_\alpha^s)_{1\leq s\leq N_{\mathrm{sim}}}$ (estimated using a kernel density estimator). Figure \ref{fig-quantiles-temp} shows the example of the station of Clermont.

\begin{figure}[H]
\centering
\includegraphics[scale=0.5]{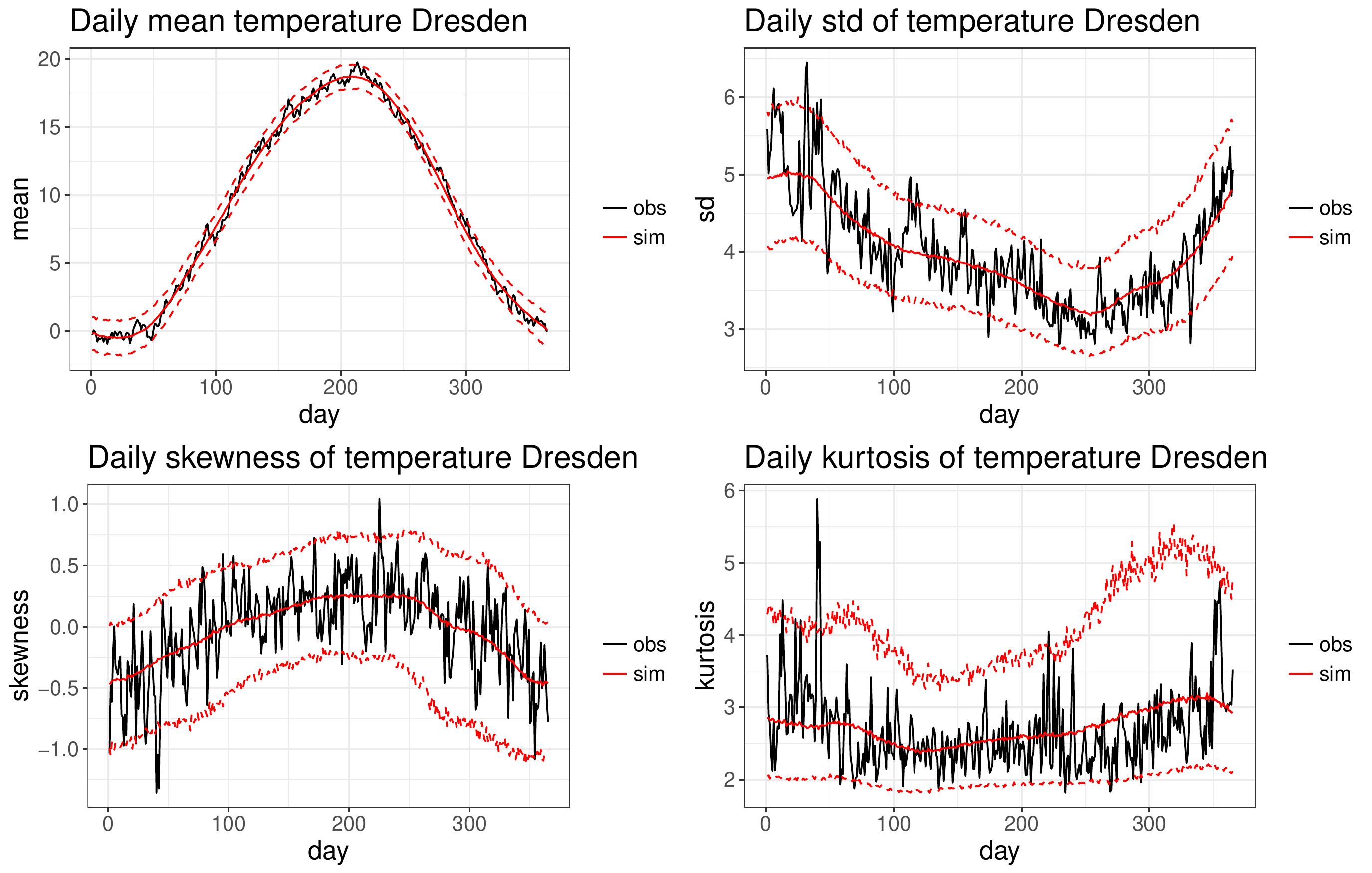}
\caption{Daily moments of temperature for Dresden: observations (black line), mean of simulations (red solid line), $2.5\%$ and $97.5\%$ quantiles of simulations (red dashed lines)}\label{fig-moments-temp}
\end{figure}

\begin{figure}[H]
\centering
\includegraphics[scale=0.5]{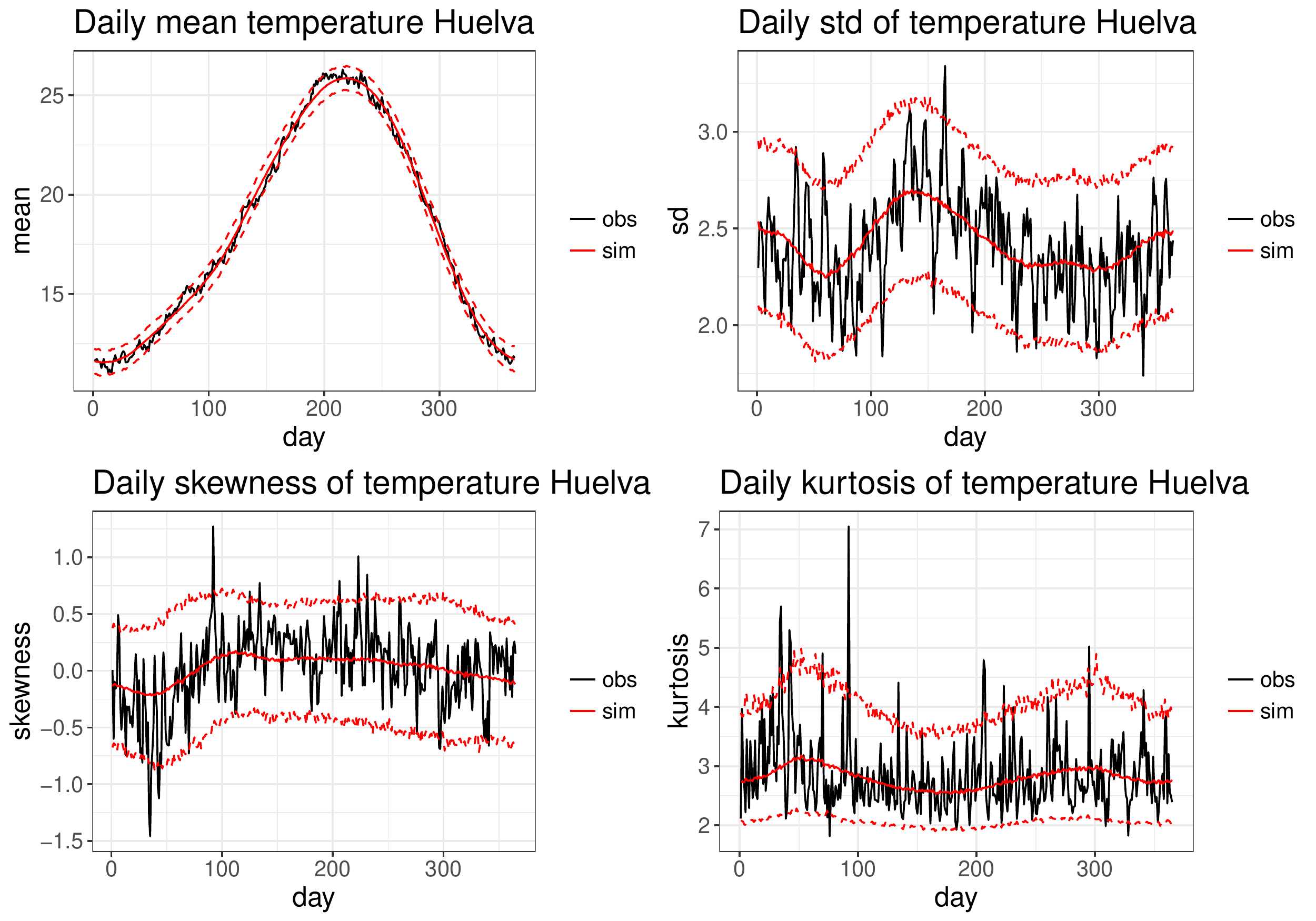}
\caption{Daily moments of temperature for Huelva}\label{fig-moments-temp-huelva}
\end{figure}

Now let us have a look at some daily statistics, beginning with daily moments of the temperature distribution. For a day of year $t\in\{1,\dots,365\}$ (e.g. January 1st), we compute the empirical mean temperature at day $t$ by averaging over the years the temperatures observed this day:
$$\bar{Y}_t^{(2)}:=\frac{1}{N_\mathrm{year}}\sum_{i=1}^{N_\mathrm{year}}Y_{t+365(i-1)}^{(2)},$$
where $N_\mathrm{year}$ is the number of observed years (here $N_{\mathrm{year}}=61$). We perform the same calculation for each simulated scenario $s\in\{1,\dots,N_\mathrm{sim}\}$, that is
$$\bar{Y}_{t,s}^{(2)}:=\frac{1}{N_\mathrm{year}}\sum_{i=1}^{N_\mathrm{year}}Y_{t+365(i-1),s}^{(2)},$$
where $Y_{t,s}^{(2)}$ is the simulated temperature at time $t$ in the $s$-th simulation. Then, using $\left(\bar{Y}_{t,s}^{(2)}\right)_{1\leq s\leq N_{\mathrm{sim}}}$, we can estimate the distribution of the mean temperature at day $t$ under the model. To be specific, we compute the mean and a $95\%$ confidence interval based on quantiles from the values $\left(\bar{Y}_{t,s}^{(2)}\right)_{1\leq s\leq N_{\mathrm{sim}}}$. Finally, the same computations are performed for all $t\in\{1,\dots,365\}$, and for the next $3$ moments: standard deviation, skewness (asymmetry coefficient) and kurtosis, as shown in Figure \ref{fig-moments-temp} for the station of Dresden and Figure \ref{fig-moments-temp-huelva} for the station of Huelva. The first daily moments are well reproduced by the model. Figure \ref{fig-moments-temp} highlights the seasonality in the variability of temperature, as the standard deviation is maximal in winter, then decreases until it reaches its minimum at the end of summer, before increasing again. The shape of this seasonality is common to all the stations we studied, except Huelva (see Figure \ref{fig-moments-temp-huelva}). This is consistent with what was observed in Figure \ref{temp_saison}. Another interesting observation is the asymmetry of the distribution of temperatures, measured by the third moment (skewness). Recall that a negative (resp. positive) skewness means that the distribution is skewed to the left (resp. right) whereas a skewness of $0$ means that the distribution is symmetric. The temperatures in Dresden clearly exhibit a seasonal behaviour in the asymmetry: the skewness is negative in winter and positive in summer. This reflects the presence of cold extremes in winter and hot extremes in summer. Using gaussian mixtures as emission distributions instead of simple gaussian distributions allows the model to reproduce this asymmetry. As Figure \ref{fig-moments-temp-huelva} shows, the station of Huelva, whose climate strongly differs from the climate of Dresden, does not exhibit the same seasonal behaviour, as the skewness curve is rather flat.

\begin{figure}[H]
\centering
\includegraphics[scale=0.5]{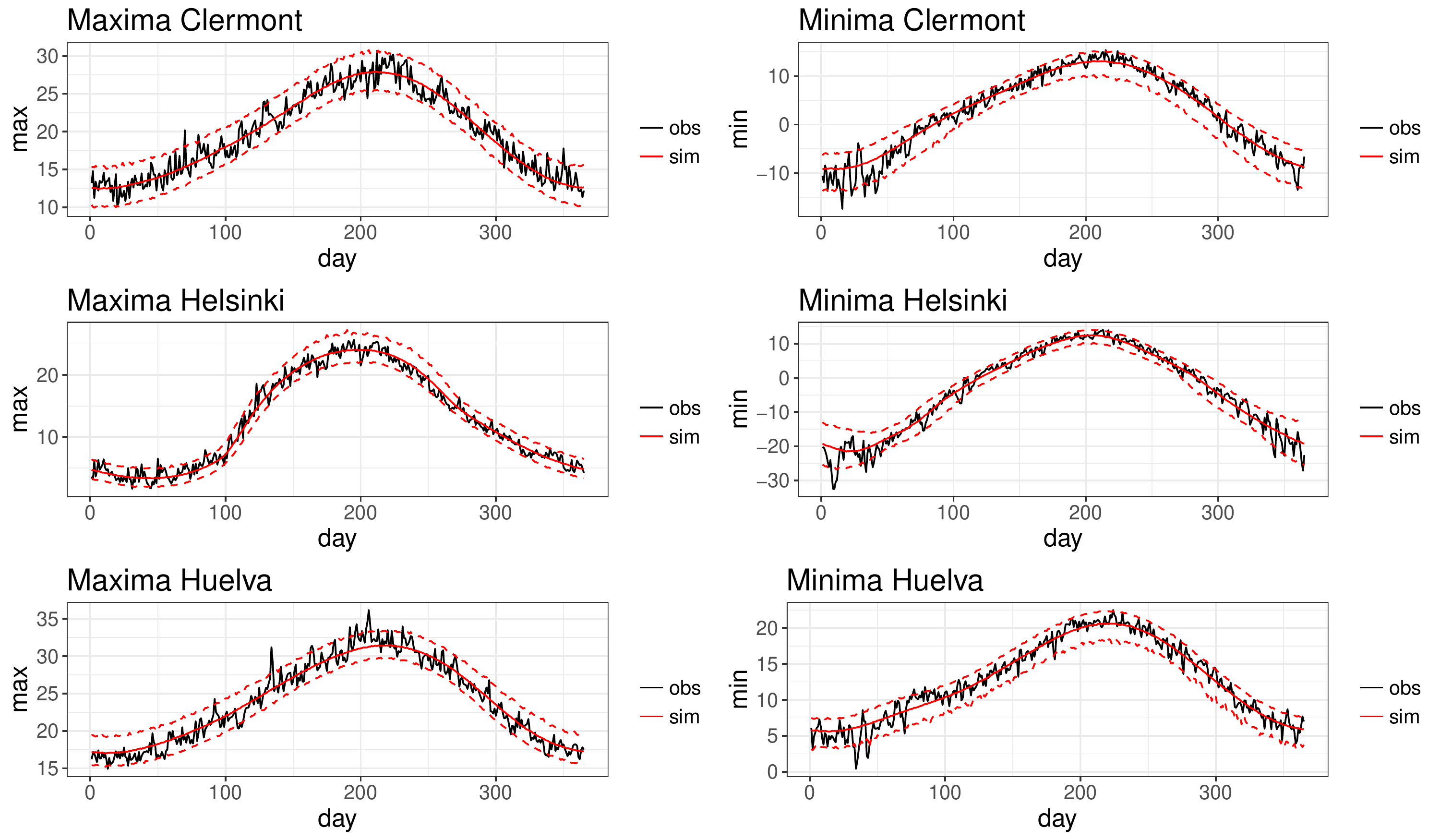}
\caption{Interannual minimum and maximum of daily mean temperature for Clermont, Helsinki and Huelva}\label{fig-max-temp}
\end{figure}

Besides the moments, it is interesting to pay attention to the interannual minimum and maximum of daily mean temperature for each day. Precisely, for a day of year $t\in\{1,\dots,365\}$, the observed maximum is $\max_{1\leq i\leq N_{\mathrm{year}}}Y_{t+365(i-1)}^{(2)}$ and we estimate the distribution of this quantity under the model using the simulations, in the same way as we did for the moments. A similar computation is performed for the daily minima. This statistic is well reproduced by the model. As an example, Figure \ref{fig-max-temp} shows the results for Clermont, Helsinki and Huelva.

\begin{figure}[H]
\centering
\includegraphics[scale=0.5]{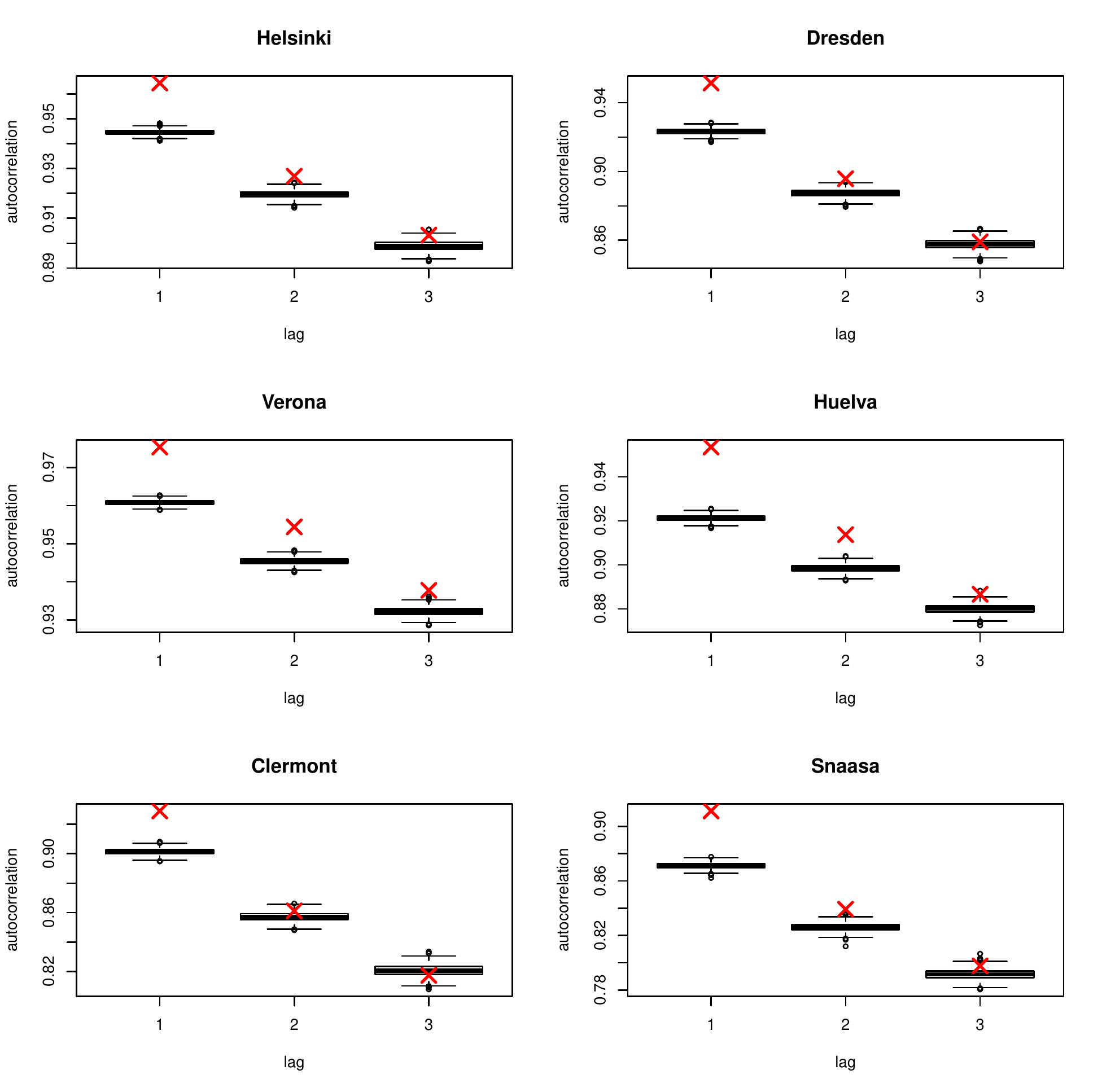}
\caption{Autocorrelation of temperatures: the red crosses are the observed values and the boxplots are related to the different simulations}\label{fig-corr-temp}
\end{figure}

The temperatures do not form an independent process, as it is strongly autocorrelated. In our model, autocorrelation is introduced through the hidden Markov chain: even though the observations are generated independently conditionally to the states, they are not independent because the state process is autocorrelated. Hence the next criterion to be considered is the empirical auto-correlation of temperature with lags $1$, $2$ and $3$ days. Figure \ref{fig-corr-temp} shows that the model slightly underestimates the autocorrelations.

\begin{figure}[H]
\centering
\includegraphics[scale=0.6]{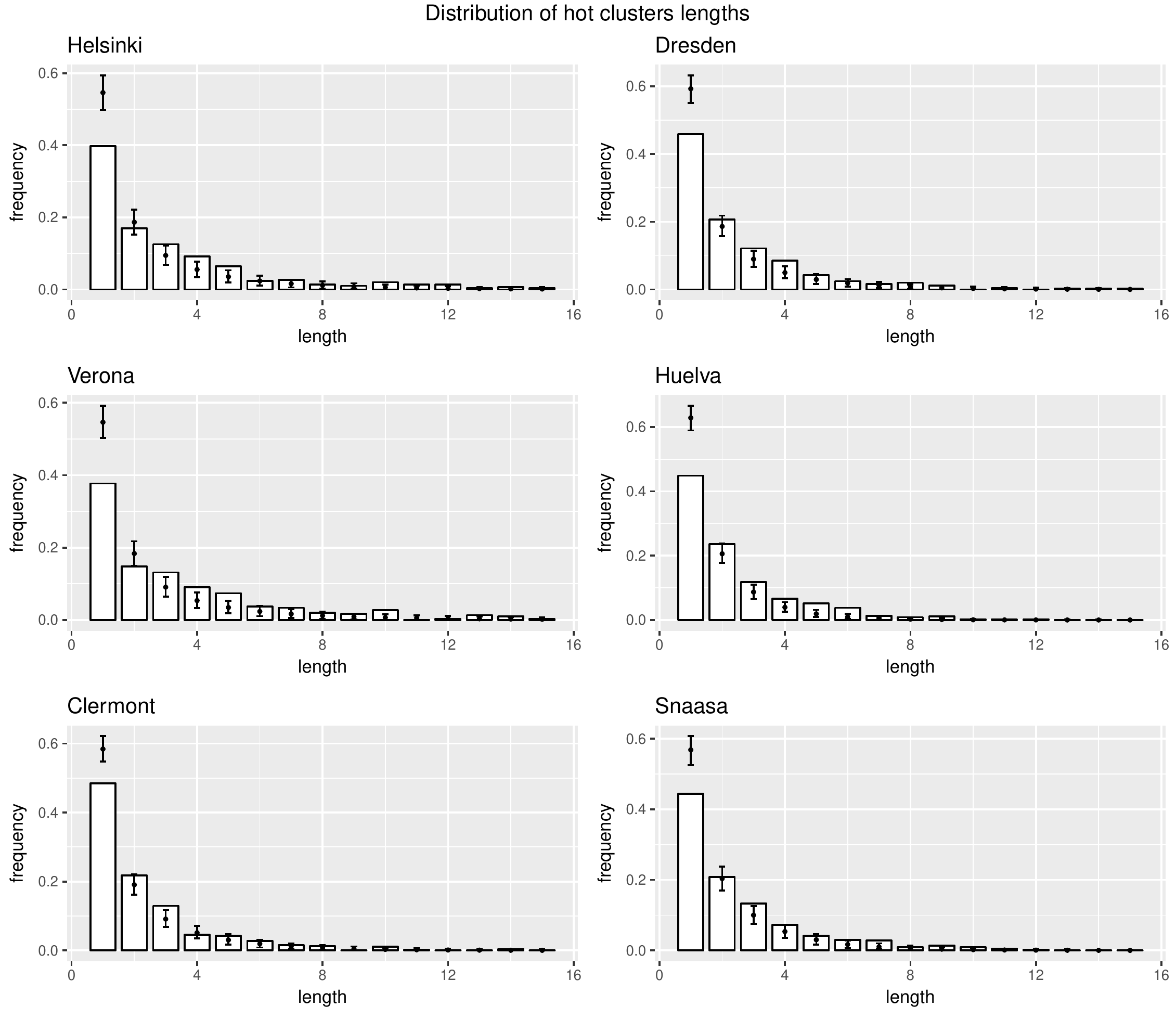}
\caption{Distribution of the lengths of the hot clusters. The threshold is the $95$-th percentile. White bars: observed values. Errorbars: $95\%$ confidence interval based on the simulations.}\label{fig-clus-chaud}
\end{figure}

\begin{figure}[H]
\centering
\includegraphics[scale=0.6]{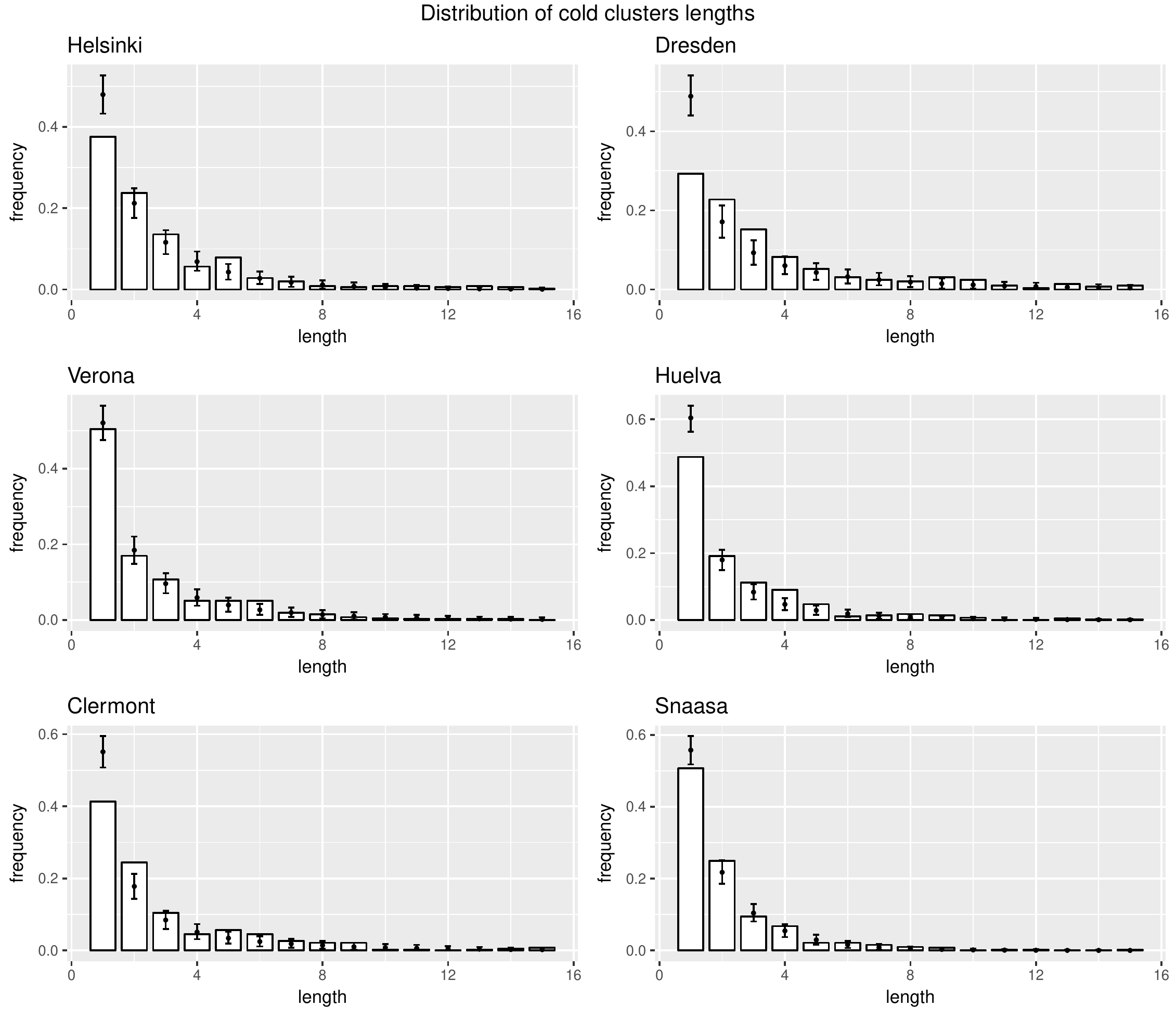}
\caption{Distribution of the lengths of the cold clusters. The threshold is the $5$-th percentile. White bars: observed values. Errorbars: $95\%$ confidence interval based on the simulations.}\label{fig-clus-froid}
\end{figure}

The last criterion that we are interested in regarding temperature is its persistence in extreme values. We fix some threshold $u$ (e.g. the $\alpha$-th quantile of temperature, with $\alpha$ close to $1$) and we consider the durations of the episodes exceeding $u$. Let us give an example. Assume that, for some $t\geq 2$, $Y_{t-1}^{(2)}\leq u$, $Y_t^{(2)}>u$, $Y_{t+1}^{(2)}>u$, $Y_{t+2}^{(2)}>u$ and $Y_{t+3}^{(2)}\leq u$. In such a case, we say that $\left(Y_{t}^{(2)},Y_{t+1}^{(2)},Y_{t+2}^{(2)} \right)$ is a \emph{hot} cluster of length $3$ because the temperature remains for $3$ days above the threshold $u$. Thus the length of a cluster is a positive integer, possibly $1$. Similarly, we can define \emph{cold} clusters by considering the times when the temperature drops below some low threshold. The results for hot clusters are shown by Figure \ref{fig-clus-chaud}. Here the threshold $u$ is the $95$-th percentile (hence it varies according to the site). Clearly, for all sites, the model produces too many clusters of length $1$, therefore not enough longer clusters. We tried various thresholds between the $90$-th and the $99$-th quantiles and the same conclusion can be drawn. Thus the persistence in extreme values is underestimated by the model. This is not surprising, considering that the autocorrelations of temperature are slightly underestimated too (see Figure \ref{fig-corr-temp}). The same issue appears when we consider cold clusters, as we can see in Figure \ref{fig-clus-froid}.

\paragraph{Precipitation}

\begin{figure}[H]
\centering
\includegraphics[scale=0.6]{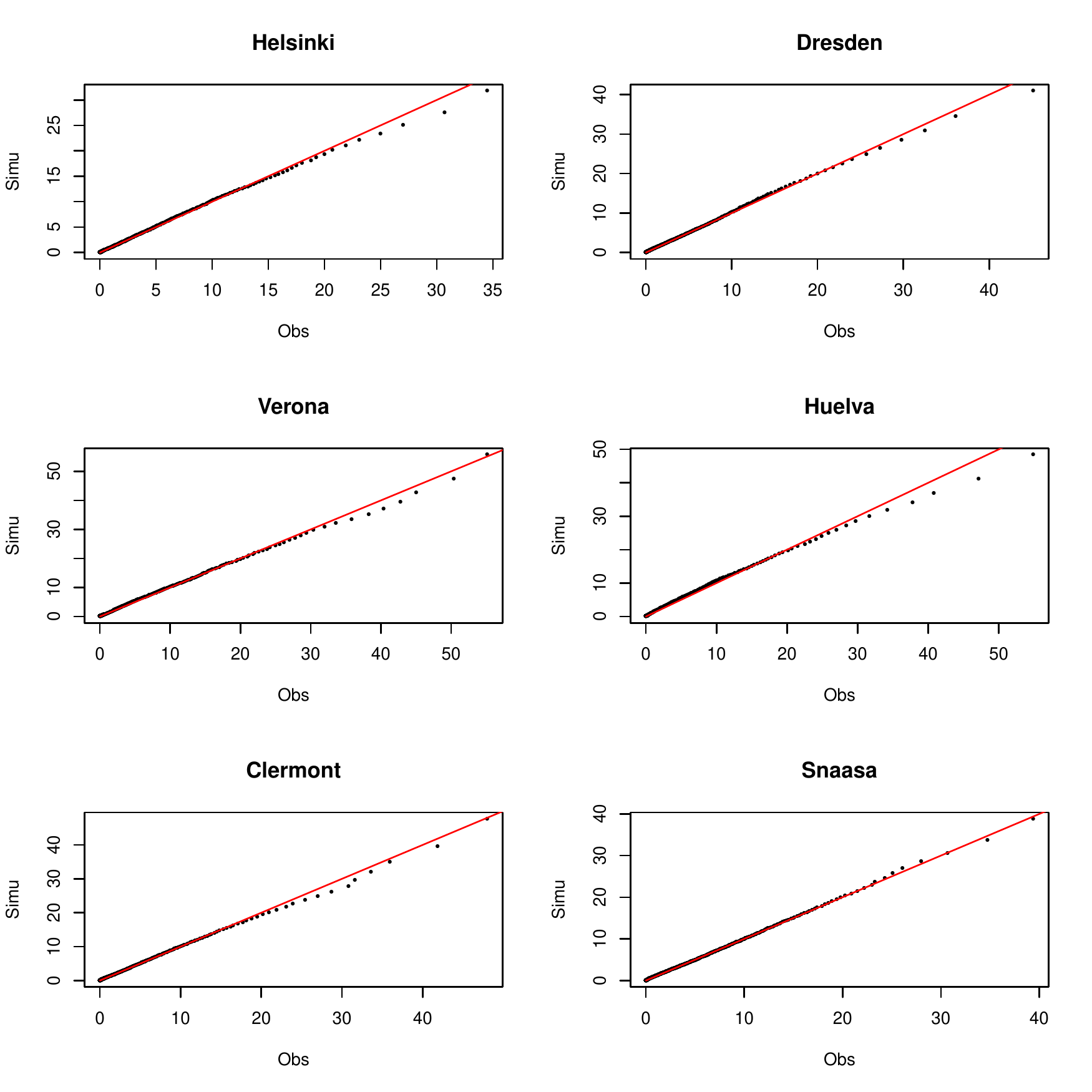}
\caption{Precipitation quantile-quantile plots.}\label{fig-qq-precip}
\end{figure}

\begin{figure}[H]
\centering
\includegraphics[scale=0.6]{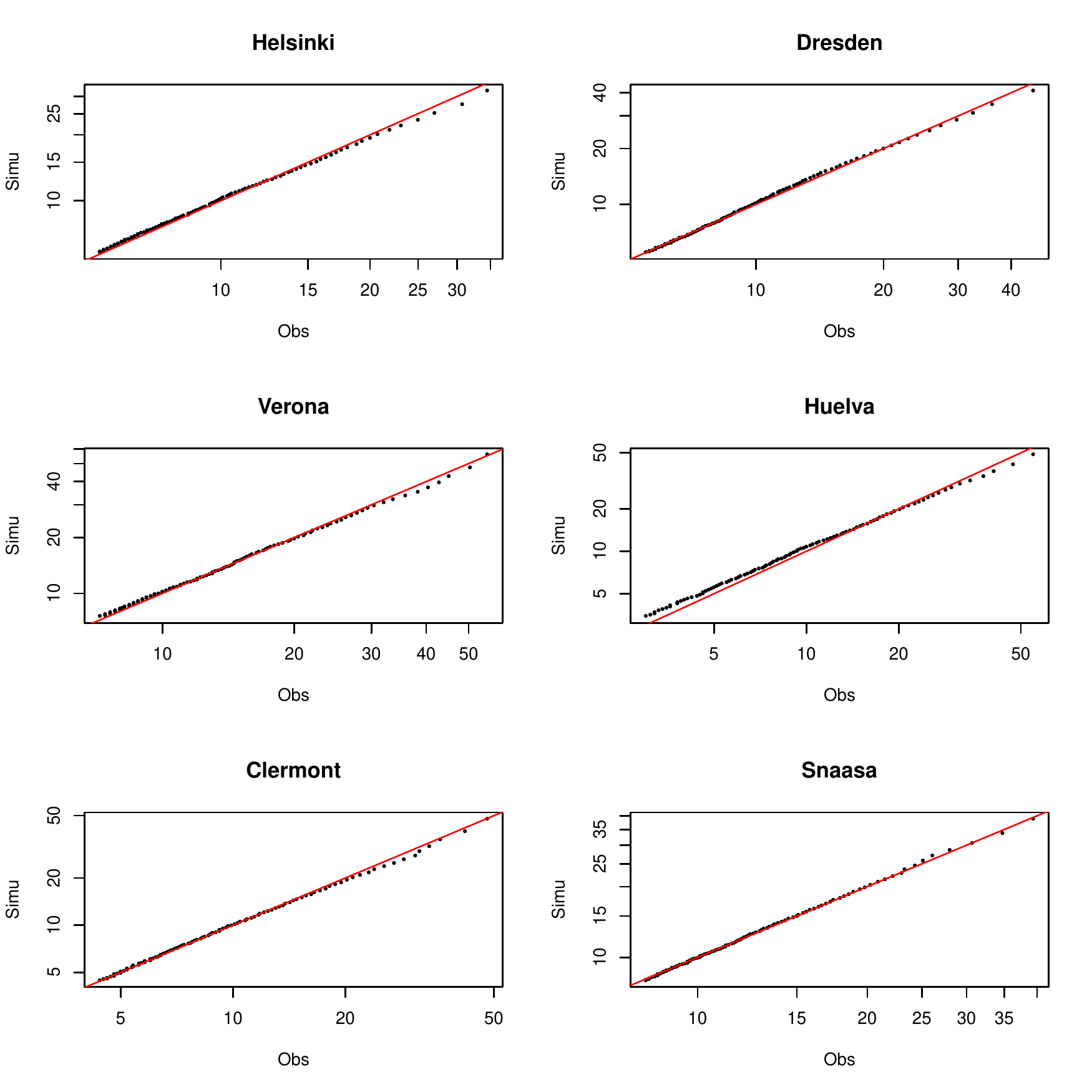}
\caption{Precipitation quantile-quantile plots, tail of distribution (logarithmic scale)}\label{fig-qq-precip-log}
\end{figure}

Figure \ref{fig-qq-precip} shows the quantile-quantile plots of precipitation, for each of the six sites. To be specific, we plotted the $\alpha$-quantiles of observed precipitations versus the corresponding simulated $\alpha$-quantiles, for $\alpha\in~\{\frac{i}{1000},\, 1\leq~i\leq~999\}$. As the distribution of precipitations is very asymmetric, we zoom-in to the $\alpha$ greater than $0.9$ (see Figure \ref{fig-qq-precip-log}). As we can see, the overall distribution of precipitations, including its tail, is well reproduced. It is also interesting to note that our model is able to simulate precipitation values that are larger than all the observed values. As an example, the maximum observd value of daily precipitation in Helsinki is $79.3$ mm, but $14$ of the $1000$ simulated trajectories include a larger value, the maximum being $110.7$ mm. This is one of the assets of model compared to models based on resampling. However, the exponential distribution being unbounded, performing a large number of simulations sometimes leads to unrealistic precipitations values.

\begin{figure}[H]
\centering
\includegraphics[scale=0.5]{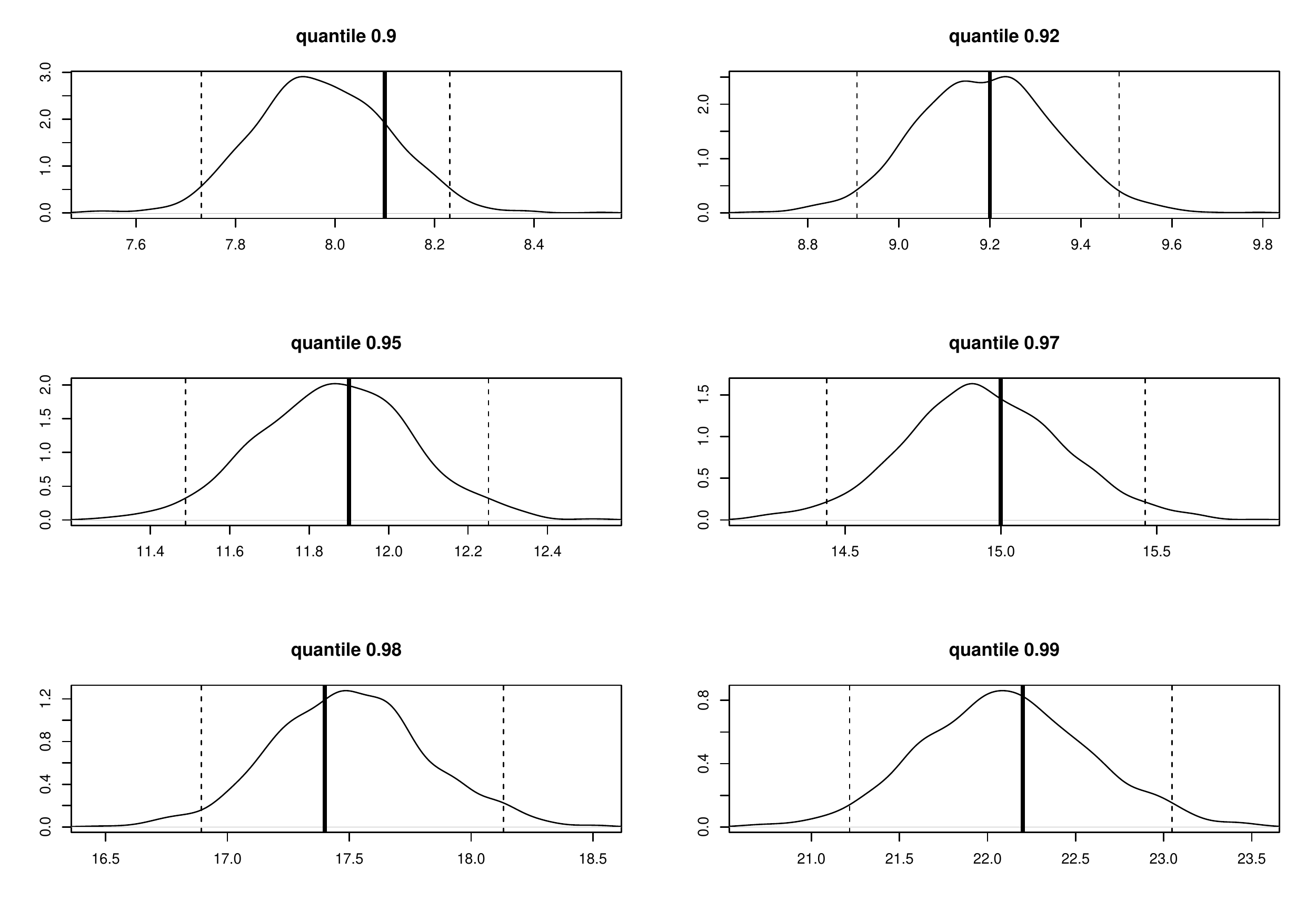}
\caption{Observed (vertical lines) superimposed to the distribution of simulated (curve) quantiles of precipitation for the station of Sn\aa sa. The dashed vertical lines are the bounds of a $95\%$ confidence interval.}\label{fig-quantiles-precip}
\end{figure}

We can also check the distributions of the simulated quantiles. As an example, Figure \ref{fig-quantiles-precip} shows the distributions of some upper quantiles for the station of Sn\aa sa, and their observed counterparts.

\begin{figure}[H]
\centering
\includegraphics[scale=0.5]{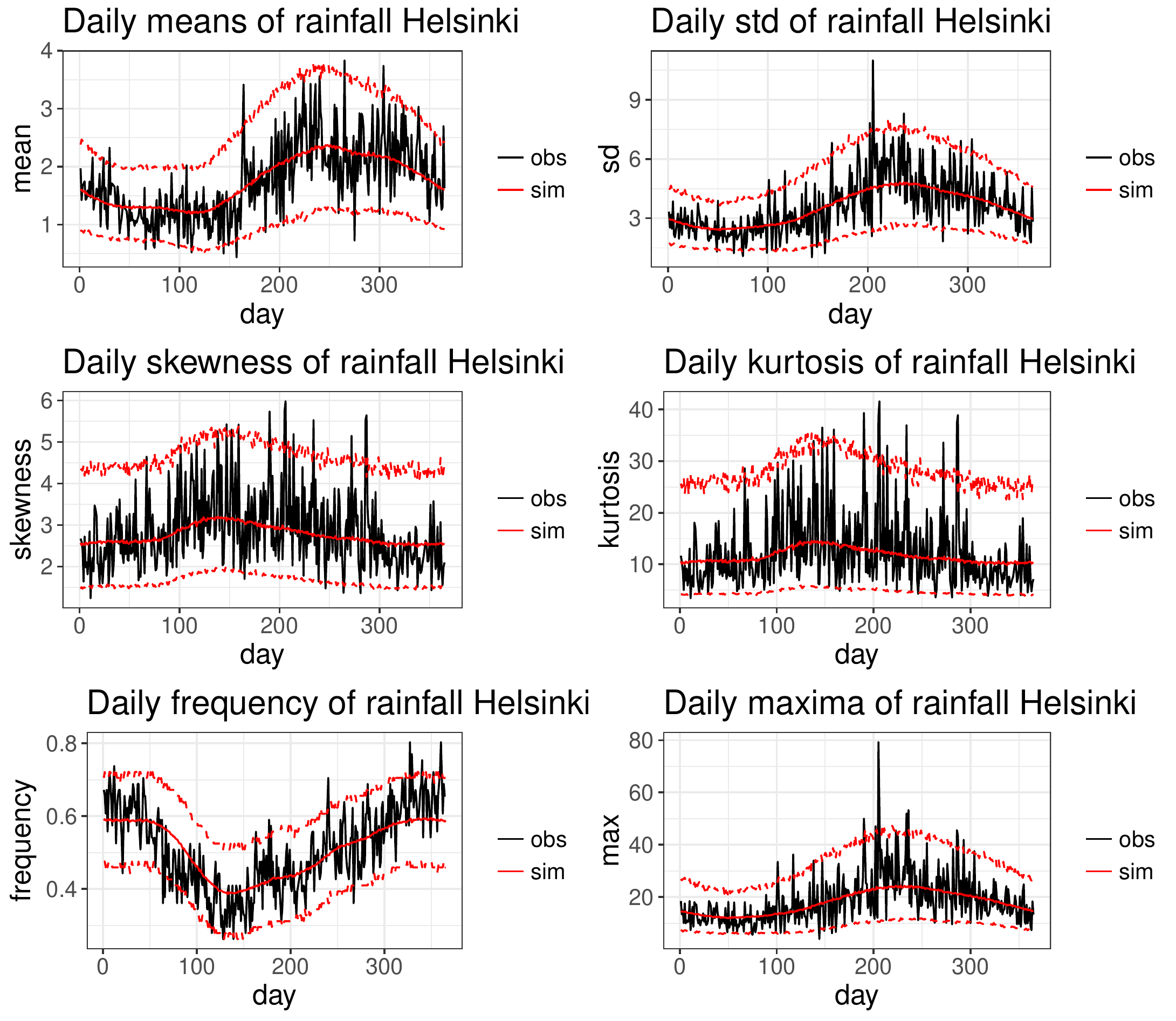}
\caption{Daily moments of precipitations}\label{fig-moments-precip}
\end{figure}

As we did for temperature, we estimate the daily moments of precipitations. We also estimate the daily frequencies of precipitations by computing, for $t\in\{1,\dots,T\}$, an estimate of $\mathbb{P}(Y_t^{(1)}>0)$ as
$$\frac{1}{N_{\mathrm{year}}}\sum_{i=1}^{N_\mathrm{year}}\mathbf{1}_{Y_{t+365(i-1)}^{(1)}>0},$$
and the maxima of daily precipitations totals:
$$\max_{1\leq i\leq N_{\mathrm{year}}}Y_{t+365(i-1)}^{(1)},\quad 1\leq t\leq 365$$
Using the simulations, we estimate the distribution of these statistics under the model. The results are presented in Figure \ref{fig-moments-precip} for the station of Helsinki, together with the first four daily moments. The seasonalities in the intensity and in the frequency of precipitations are well reproduced by the model.

\begin{figure}[H]
\centering
\includegraphics[scale=0.5]{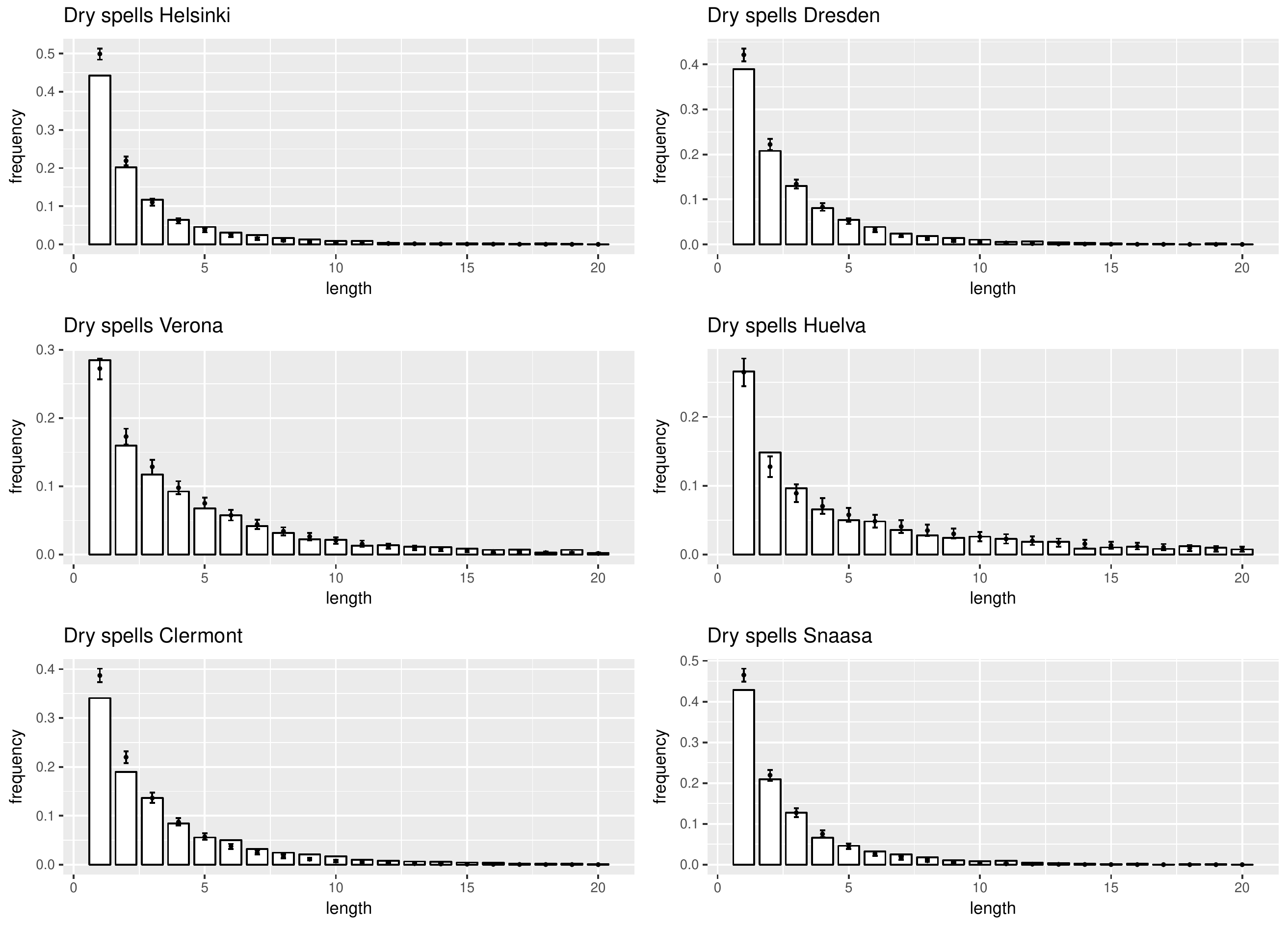}
\caption{Distribution of the length of dry spells. White bars: observed values. Errorbars: $95\%$ confidence interval based on the simulations.}\label{fig-dry-spells}
\end{figure}

\begin{figure}[H]
\centering
\includegraphics[scale=0.5]{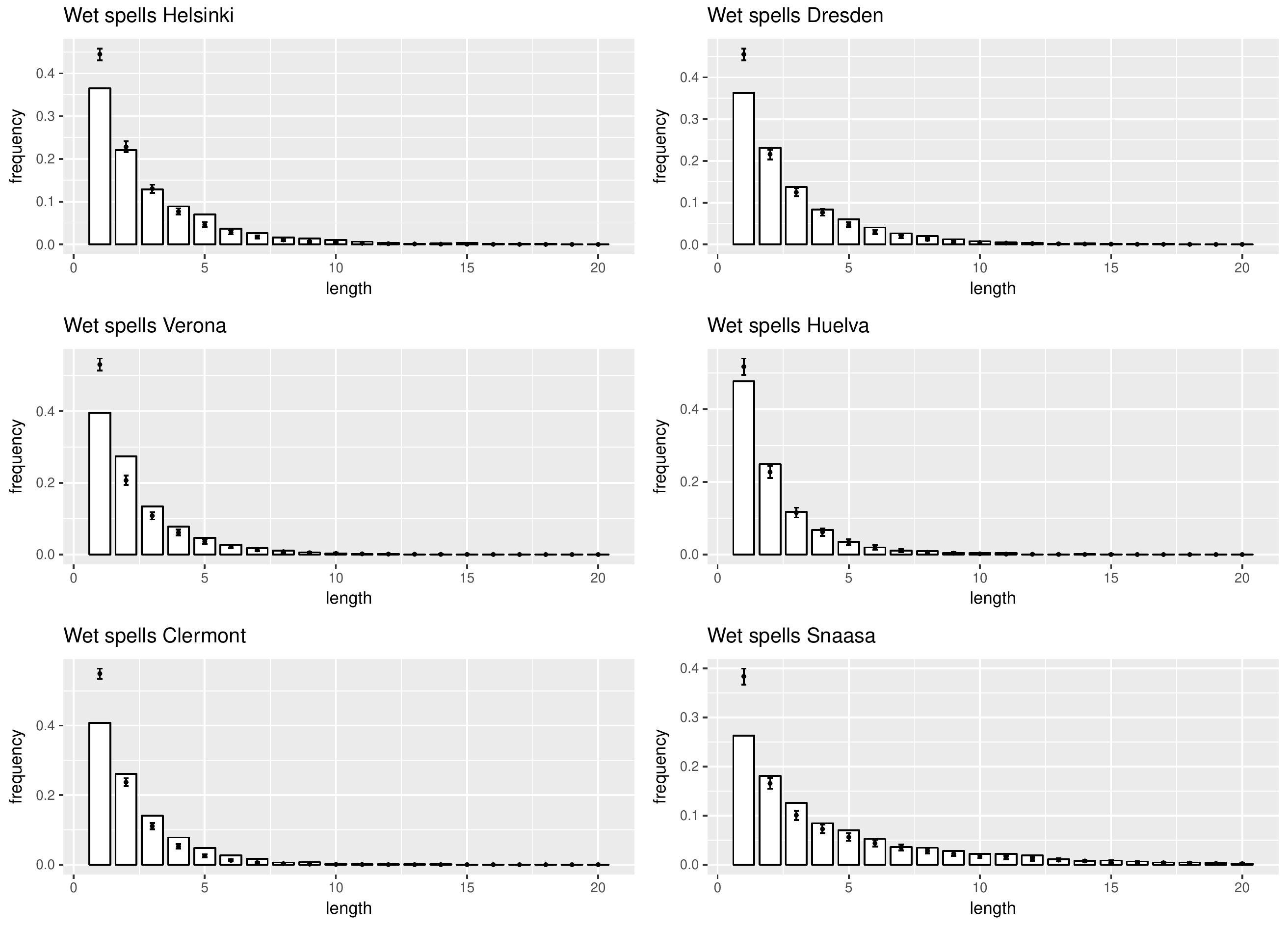}
\caption{Distribution of the length of wet spells. White bars: observed values. Errorbars: $95\%$ confidence interval based on the simulations.}\label{fig-wet-spells}
\end{figure}

A dry spell is a period of time during which it does not rain. When there are $d$ consecutive days without rain and non-zero precipitations on the $(d+1)$-th day, this constitutes a dry spell of length $d$. Similarly, we define wet spells as consecutive days with non-zero precipitations. Figures \ref{fig-dry-spells} and \ref{fig-wet-spells} show the observed and simulated distributions of dry and wet spells. The lengths of dry spells are quite well reproduced by the model but for some stations (e.g. Clermont), the model clearly underestimates the number of wet spells longer than one day.

\begin{figure}[H]
\centering
\includegraphics[scale=0.5]{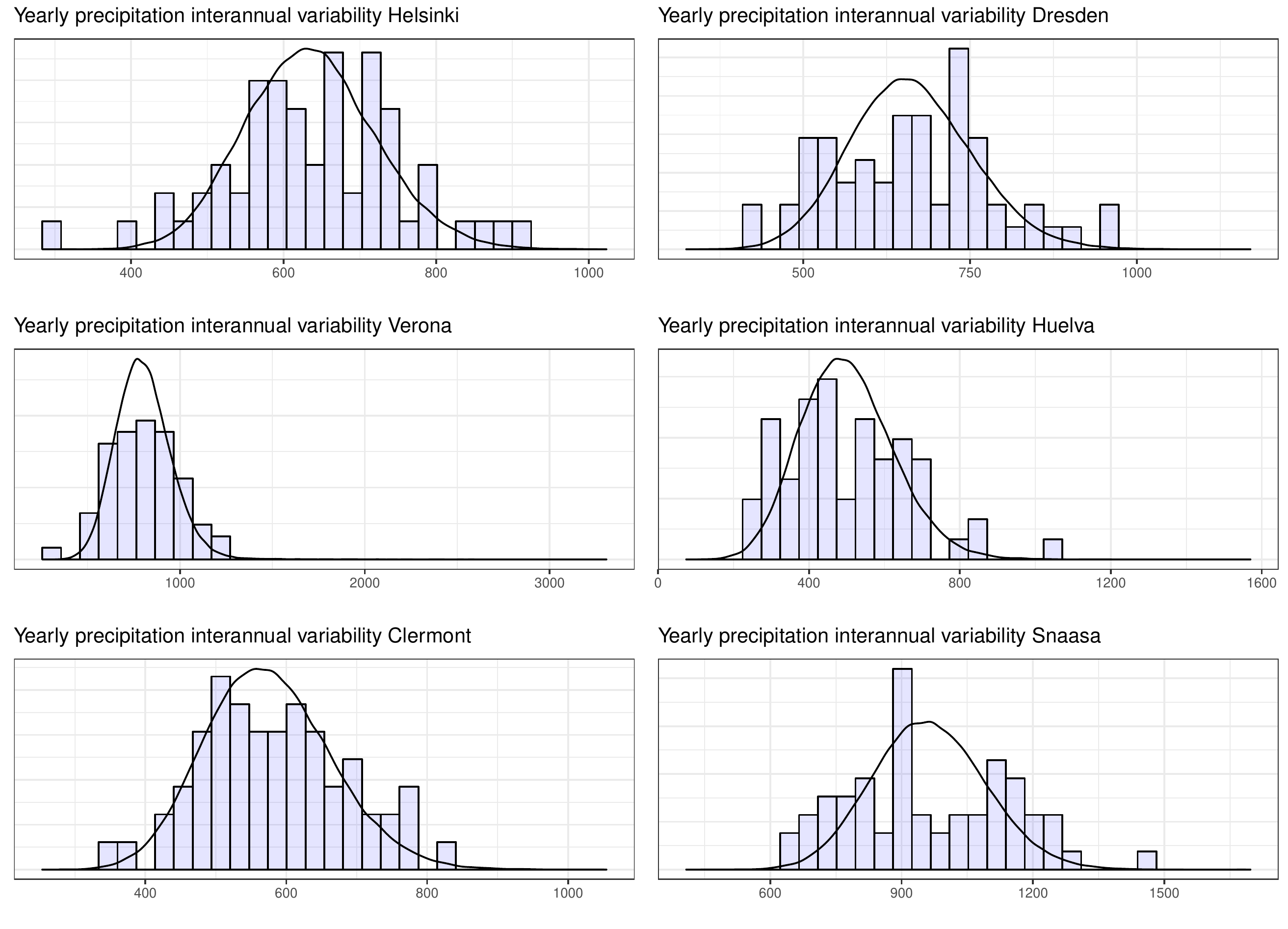}
\caption{Interannual variability of yearly precipitations}\label{fig-interannual}
\end{figure}

\begin{figure}[H]
\centering
\includegraphics[scale=0.5]{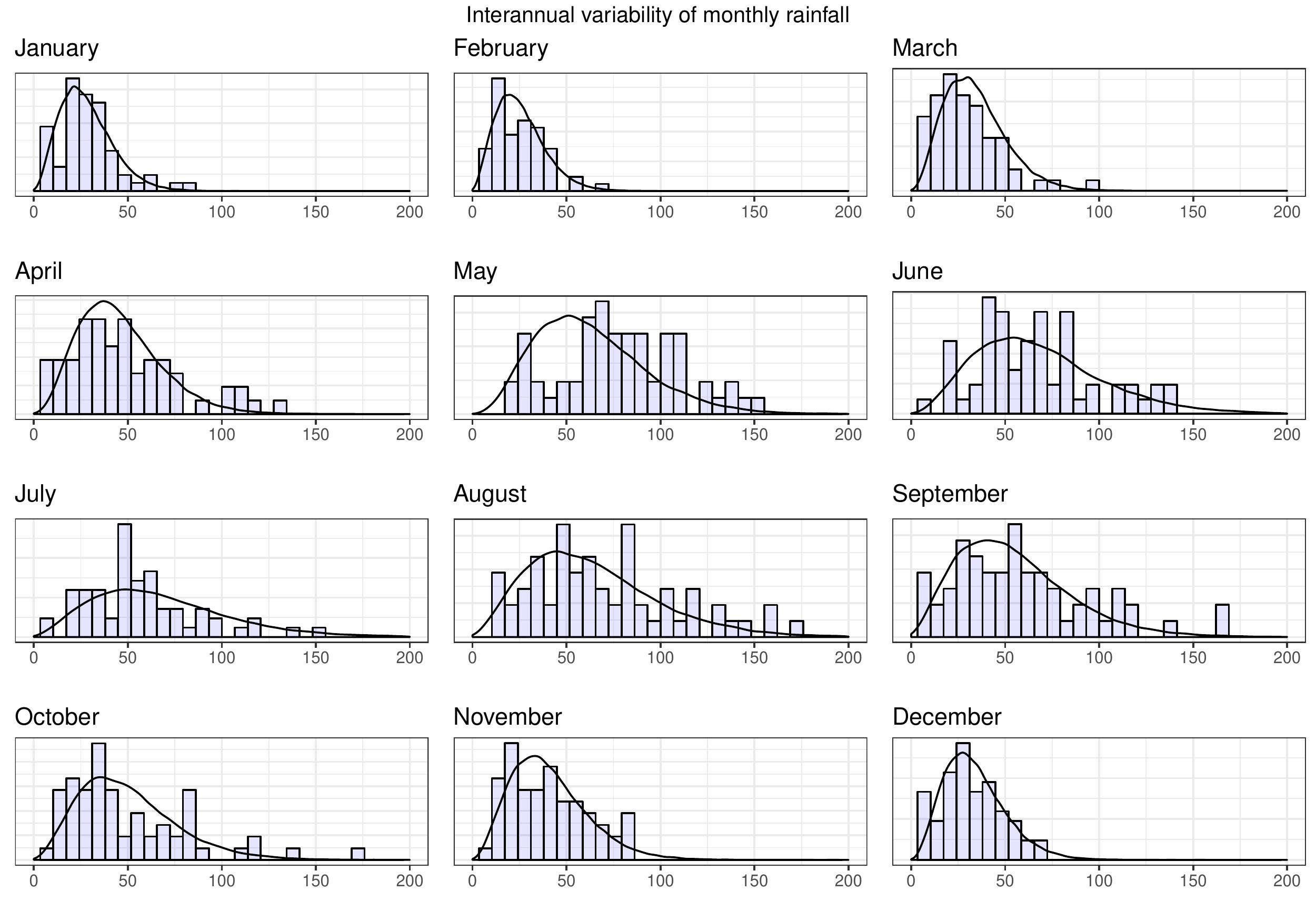}
\caption{Interannual variability of monthly precipitations}\label{fig-interannual-month}
\end{figure}

Stochastic precipitations generators often underestimate the interannual variability of precipitations \citep{katz1998}. Thus we focus on yearly rainfall and we look at its interannual variability. The histograms in Figure \ref{fig-interannual} are the observed distributions of yearly precipitations (thus each histogram has been computed with $61$ observations). The lines are the kernel density estimations of simulated yearly precipitations. We have performed the same computations with monthly precipitations (see Figure \ref{fig-interannual-month} for the station of Clermont). Our model does not underestimate interannual variability, as it is able to generate rainy as well as dry years or months.

\paragraph{Temperature and precipitation coupling}

\begin{figure}[H]
\centering
\includegraphics[scale=0.5]{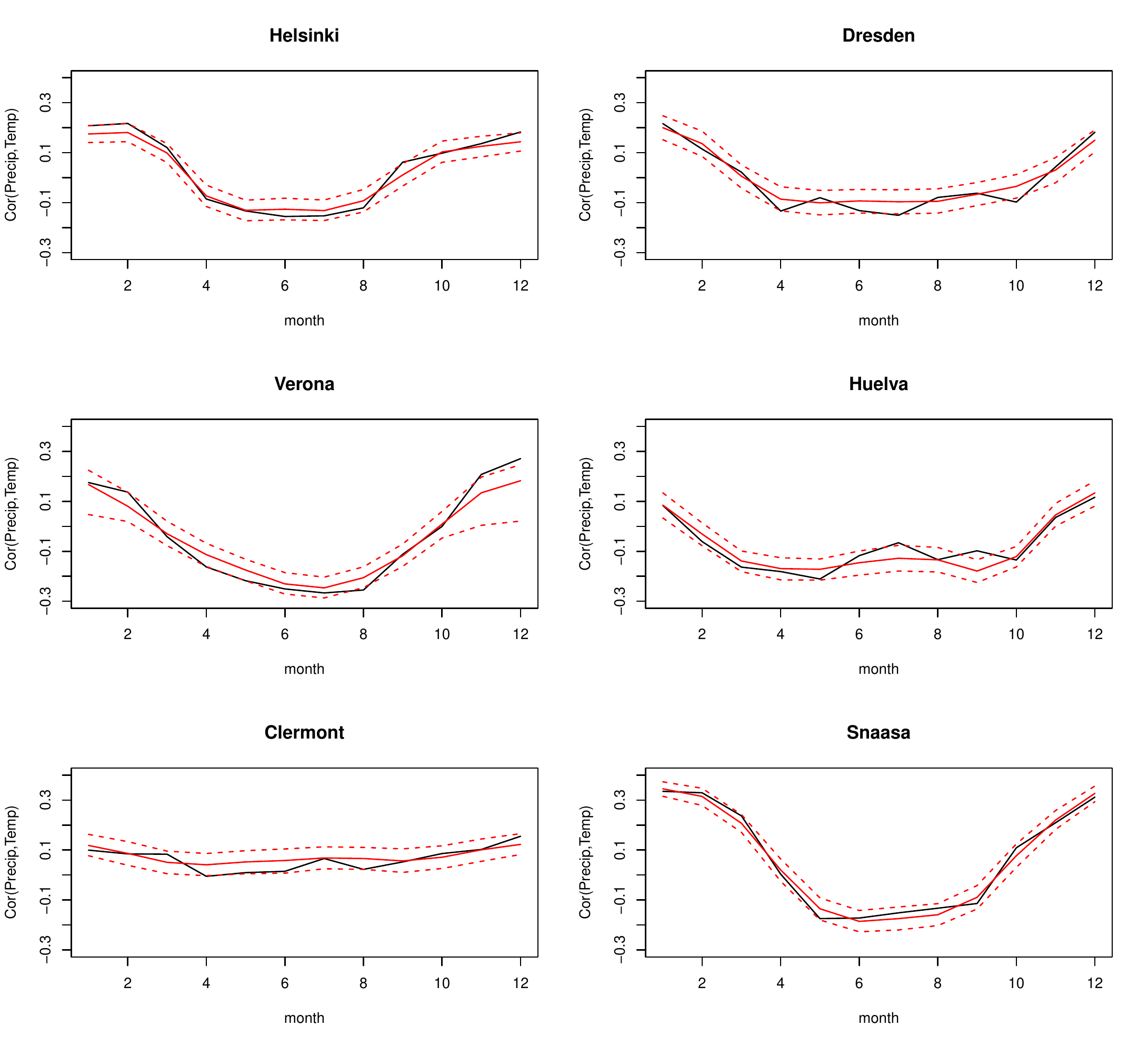}
\caption{Monthly correlations between temperature and precipitations. Black solid line: observed. Red solid line: mean of simulations. Red dashed lines: $95\%$ confidence interval from simulations.}\label{fig-cor-mois}
\end{figure}

At this stage, we have assessed the performance of our model for temperature and precipitations separately. We shall now concentrate on the relationship between these two variables, as they are not independent. Figure \ref{fig-cor-mois} shows that the model provides realistic monthly correlations between temperature and precipitations.

However, a better representation of the relationship between temperature and precipitations can be obtained. For example, the probability of observing precipitations varies with temperature:  for $u,v\in\mathbb{R}$, in general,
$$\mathbb{P}\left(Y_t^{(1)}>0\mid Y_t^{(2)}=u\right)\neq \mathbb{P}\left(Y_t^{(1)}>0\mid Y_t^{(2)}=v\right).$$
Similarly, the expected value of non-zero precipitation depends on temperature: for $u,v\in\mathbb{R}$, in general,
$$\mathbb{E}\left[Y_t^{(1)}\mid Y_t^{(1)}>0,Y_t^{(2)}=u\right]\neq \mathbb{E}\left[Y_t^{(1)}\mid Y_t^{(1)}>0,Y_t^{(2)}=v\right].$$
Note that the quantities $\mathbb{P}\left(Y_t^{(1)}>0\mid Y_t^{(2)}=u\right)$ and $\mathbb{E}\left[Y_t^{(1)}\mid Y_t^{(1)}>0,Y_t^{(2)}=u\right]$ depend not only on $u$ but also on $t$ as the process $\left(Y_t^{(1)},Y_t^{(2)}\right)_{t\geq 1}$ is not stationary.

Let $\mathcal{K}$ be the gaussian kernel defined by $\mathcal{K}(x) = \exp\left(-\frac{x^2}{2}\right)$. For $y\in\mathbb{R}$ and a bandwidth $h>0$, we consider the following statistics.

$$r(y) := \frac{\sum_{t=1}^n\mathcal{K}\left(\frac{Y_t^{(2)}-y}{h}\right)\mathbf{1}_{Y_t^{(1)}>0}}{\sum_{t=1}^n\mathcal{K}\left(\frac{Y_t^{(2)}-y}{h}\right)}$$

$$R(y) := \frac{\sum_{t=1}^n\mathcal{K}\left(\frac{Y_t^{(2)}-y}{h}\right)Y_t^{(1)}}{\sum_{t=1}^n\mathcal{K}\left(\frac{Y_t^{(2)}-y}{h}\right)\mathbf{1}_{Y_t^{(1)}>0}}$$

\begin{figure}[H]
\centering
\includegraphics[scale=0.6]{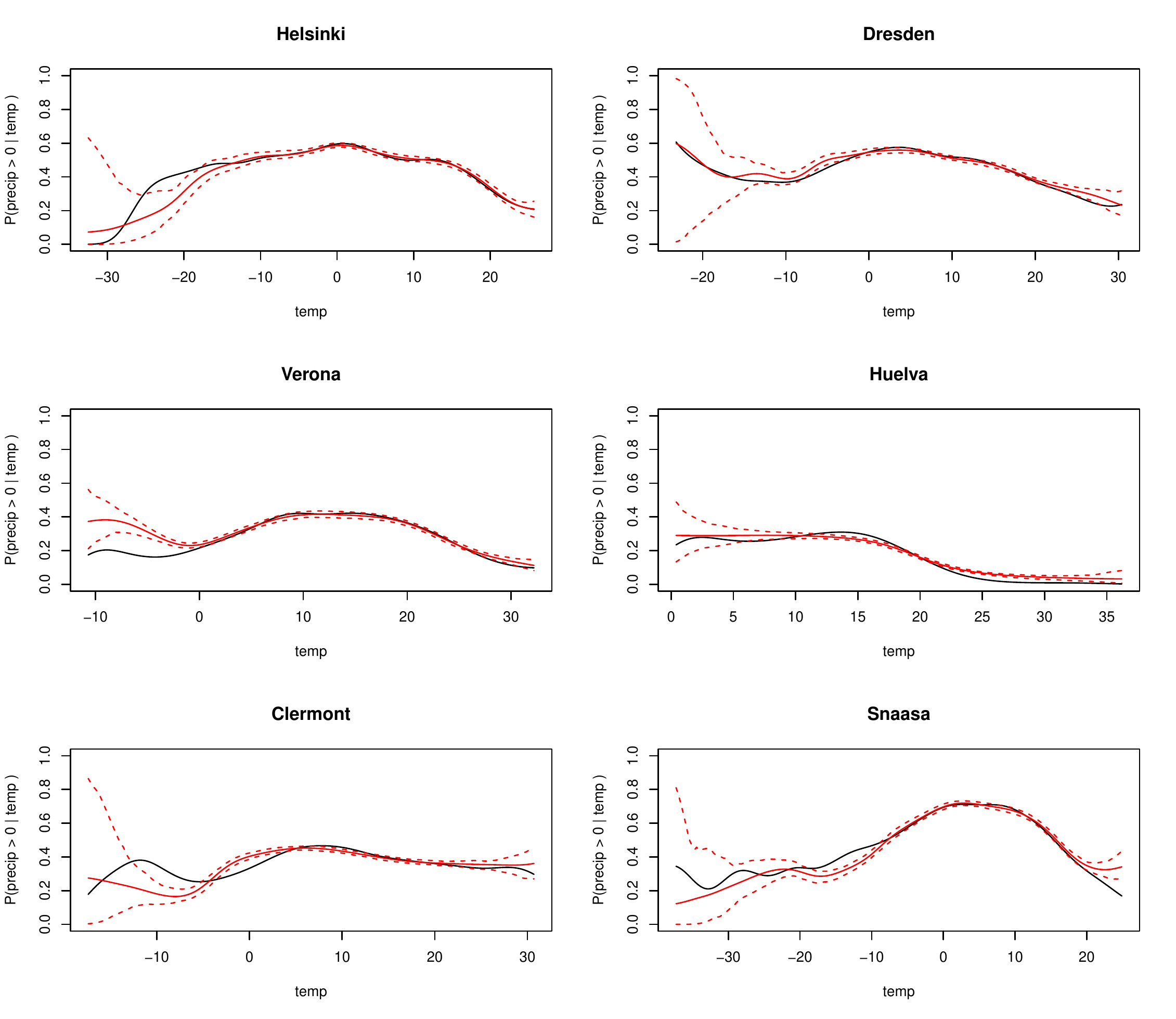}
\caption{Graph of the function $y\mapsto r(y)$. Solid black curve: computation on the observations. Solid red curve: mean of the simulated values. Dashed red curves: $95\%$ confidence interval based on simulations. }\label{fig-pcond}
\end{figure}

\begin{figure}[H]
\centering
\includegraphics[scale=0.5]{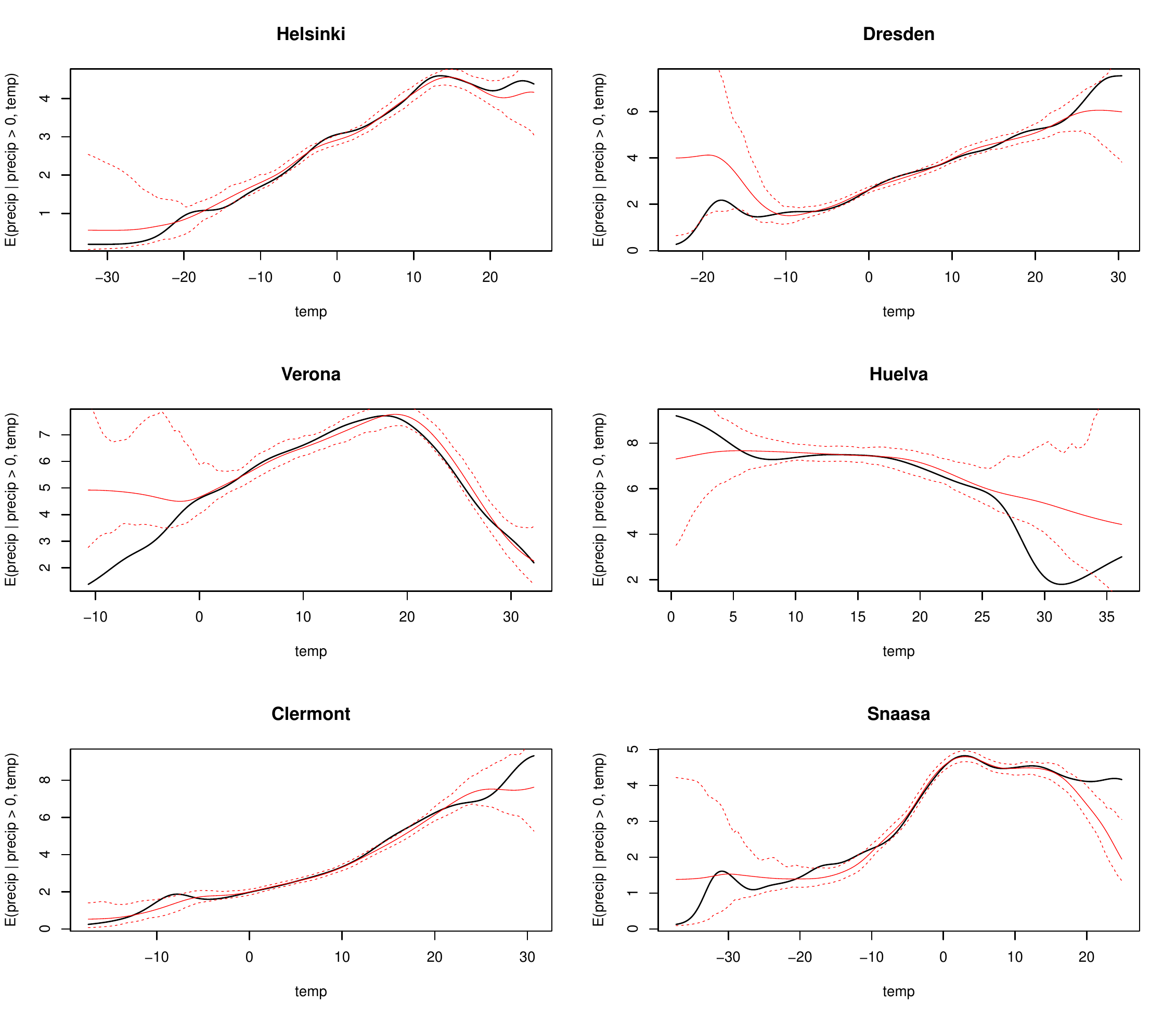}
\caption{Graph of the function $y\mapsto R(y)$. Solid black curve: computation on the observations. Solid red curve: mean of the simulated values. Dashed red curves: $95\%$ confidence interval based on simulations.}\label{fig-econd}
\end{figure}

If we had a sample $\left(Y_t^{(1)},Y_t^{(2)}\right)_{1\leq t\leq n}$ of i.i.d. copies of $\left(Y^{(1)},Y^{(2)}\right)$, then $r(y)$ would be an estimator of $\mathbb{P}\left(Y^{(1)}>0\mid Y^{(2)}=y\right)$ and $R(y)$ would be an estimator of $\mathbb{E}\left[Y^{(1)}\mid Y^{(1)}>0, Y^{(2)}=y\right]$. Although this is not the case, these statistics still provide some information on the dependence between precipitation occurrence and temperature and it is interesting to see how the model behaves with respect to $r(y)$ and $R(y)$. Therefore, the functions $y\mapsto r(y)$ and $y\mapsto R(y)$ are computed from the observations and from each of the $1000$ bivariate simulations. Figures \ref{fig-pcond} and \ref{fig-econd} show the results for the six stations (with $h=2$). Here again, the model is performing well with regard to these statistics.

For $t\in\{1,\dots n\}$, we denote by $\bar{t}\in\{1,\dots,365\}$ its representative modulo $365$, that is the day of year. For $t,s\in\{1,\dots,n\}$, let
$$\| t-s\|:=\min (|\bar{t}-\bar{s}|,365 - |\bar{t}-\bar{s}|)$$
be the cyclic distance between days $t$ and $s$, that is the number of days between the corresponding days of year. For $h_1,h_2>0$, $y\in\mathbb{R}$ and $t\in\{1,\dots, 365\}$, let us define
$$r(t,y) := \frac{\sum_{s=1}^n\mathcal{K}\left(\frac{\| t-s\|}{h_1}\right)\mathcal{K}\left(\frac{y-Y^{(2)}_s}{h_2}\right)\mathbf{1}_{Y_s^{(1)}>0}}{\sum_{s=1}^n\mathcal{K}\left(\frac{\| t-s\|}{h_1}\right)\mathcal{K}\left(\frac{y-Y^{(2)}_s}{h_2}\right)},$$
and
$$R(t,y) := \frac{\sum_{s=1}^n\mathcal{K}\left(\frac{\| t-s\|}{h_1}\right)\mathcal{K}\left(\frac{y-Y^{(2)}_s}{h_2}\right)Y_s^{(1)}}{\sum_{s=1}^n\mathcal{K}\left(\frac{\| t-s\|}{h_1}\right)\mathcal{K}\left(\frac{y-Y^{(2)}_s}{h_2}\right)\mathbf{1}_{Y_s^{(1)}>0}}.$$

\begin{figure}[H]
\centering
\includegraphics[scale=0.5]{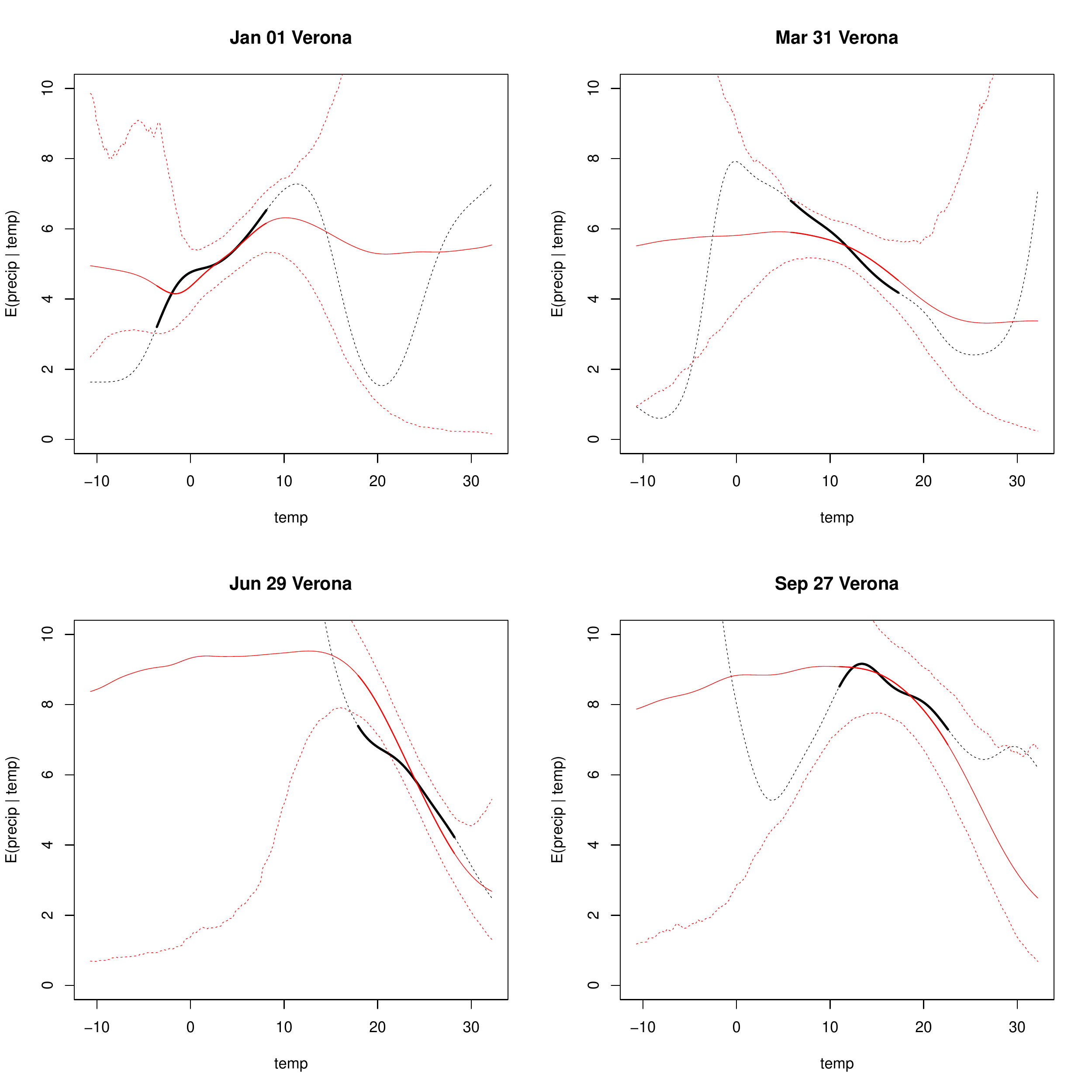}
\caption{Graphs of the function $y\mapsto r(t,y)$ for different values of $t$ (day of year) in Verona. Solid black curve: computation on the observations. Solid red curve: mean of the simulated values. Dashed red curves: $95\%$ confidence interval based on simulations. The dashed black curve is an extrapolation of $y\mapsto r(t,y)$ outside the observed range of temperature at day of year $t$: although the computation is feasible, it is not relevant.}\label{fig-econd-verona}
\end{figure}

\begin{figure}[H]
\centering
\includegraphics[scale=0.5]{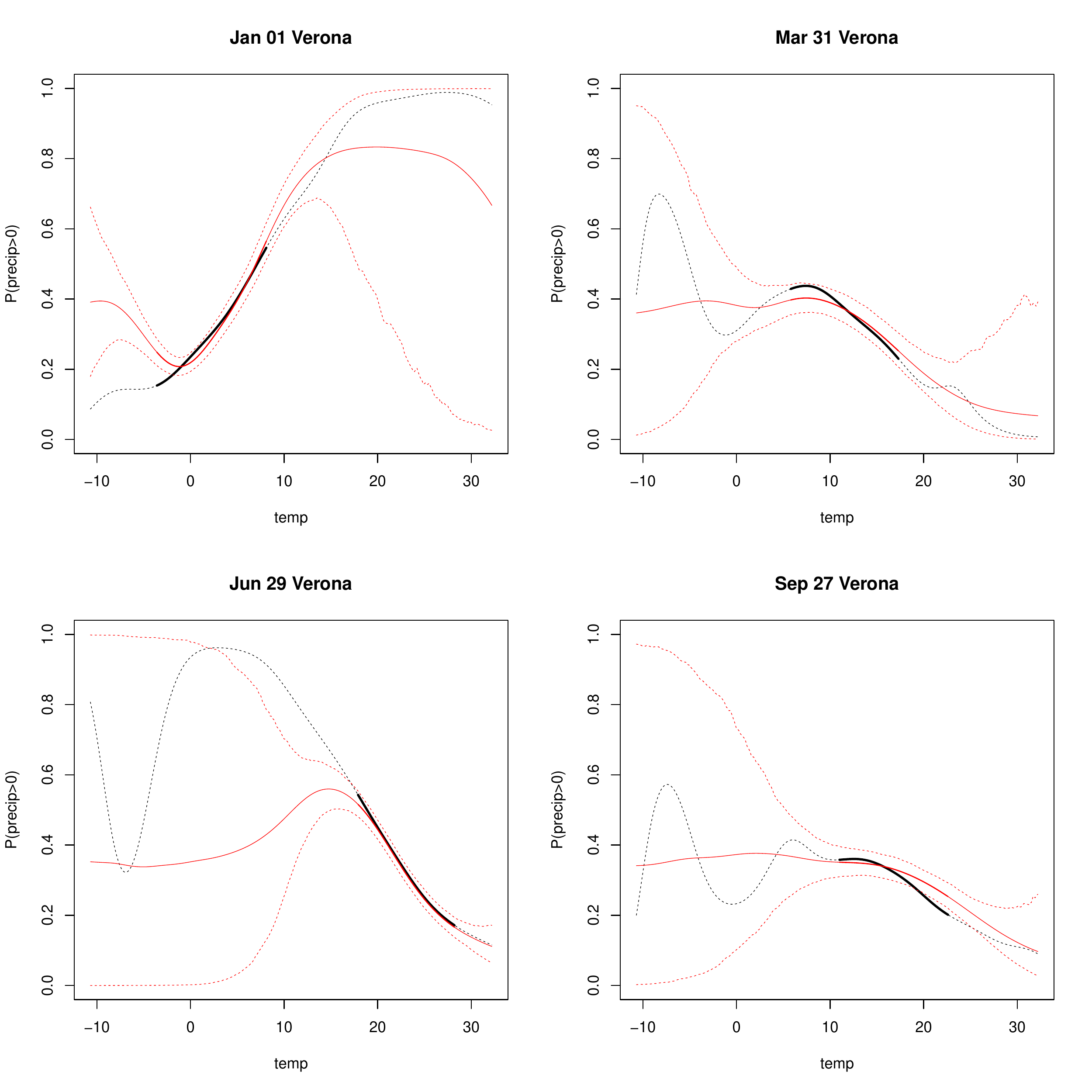}
\caption{Graphs of the function $y\mapsto R(t,y)$ for different values of $t$ (day of year) in Verona. Solid black curve: computation on the observations. Solid red curve: mean of the simulated values. Dashed red curves: $95\%$ confidence interval based on simulations. The dashed black curve is an extrapolation of $y\mapsto r(t,y)$ outside the observed range of temperature at day of year $t$: although the computation is feasible, it is not relevant.}\label{fig-pcond-verona}
\end{figure}

Then $r(t,y)$ is a proxy for $\mathbb{P}\left(Y_t^{(1)}>0\mid Y_t^{(2)}=y\right)$ and $R(t,y)$ is a proxy for $\mathbb{E}\left[Y_t^{(1)}\mid Y_t^{(1)}>0,Y_t^{(2)}=u\right]$. Hence we use these statistics as validation criteria. Figures \ref{fig-econd-verona} and \ref{fig-pcond-verona} show the results for the station of Verona for four different days of year. Results are in general satisfying and demonstrate a realistic coupling between the two variables.

\section{Conclusion}

We introduced a seasonal hidden Markov model for the joint modeling of daily temperature and precipitations. The non-homogeneity of the underlying Markov chain allows the model to account for the complex seasonal features of these weather variables, as well as climate change, in a unified framework, without resorting to pre-processing the data or fitting multiple models. Our model can be used as a stochastic weather generator, as it can quickly generate realistic synthetic time series of temperature and precipitations, at a given site. Considering many criteria of interest, we showed that these simulations closely reproduce the behaviour of the data, be it the marginal distributions of the two variables or their dependence relationships. Furthermore, we showed that investigating the estimated parameters of the model leads to giving \emph{a posteriori} a physical intepretation to the hidden states, thus avoiding the pitfall of a \emph{black box} model. We also proved the robustness of our model by testing it on different sites with various climates.\\ 

Several extensions of our model can be considered and be the subject of future works. First, we noted that in many cases, it fails to reproduce correctly the extreme heat or cold episodes, and the dry and rainy spells. This flaw can be caused by a lack of autocorrelation and could be adressed by adding autoregression in the process. Using our notations, the distribution of $Y_t$ could depend on $t$, $Y_{t-1}$ and $X_t$ instead of just being a function of $t$ and $X_t$. Then, extreme values of temperature and precipitations can be investigated more closely. We did not focus on this particular point but some applications need a fine modeling of extremes. To this aim, it may be necessary to choose other emission distributions, even though we showed that the upper quantiles of temperature and precipitations were well reproduced. In order to apply the model to sites where there is a sensible trend in the distibution of precipitations, it would have to be modified. Finally, the structure of our model can easily be extended to more variables (e.g. wind speed), the main difficulty being the choice of the emission distributions.

\bibliographystyle{plainnat}
\bibliography{bibli}

\end{document}